\documentclass[12pt,preprint]{aastex}
\usepackage{amsmath}
\usepackage{amssymb}
\usepackage{epsfig}

\newcommand{\cms}{\,{\rm cm$^{-2}$}\,}
\newcommand{\cmc}{\,{\rm cm$^{-3}$}\,}
\newcommand{\kms}{\,{\rm km\,s$^{-1}$}\,}

\newcommand{\etal}{{ et~al.~}}
\newcommand{\expunit}{\,{\rm photons\,s$^{-1}$\,cm$^{-2}$\,arcsec$^{-2}$}\,}

\newcommand{\ergs}{\,{\rm erg\,s$^{-1}$}\,}
\newcommand{\ergscm}{\,{\rm erg\,s$^{-1}$\,cm$^{-2}$}\,}
\newcommand{\Ms}{M_\odot}
\newcommand{\Zs}{Z_\odot}

\shorttitle{Chandra Observations of NGC 4438}

\begin{document}


\title{Chandra Observations of NGC 4438: An Environmentally Damaged
Galaxy in the Virgo Cluster}

\author{Marie E. Machacek, Christine Jones, and William R. Forman}
\affil{Harvard-Smithsonian Center for Astrophysics, MS67, \\ 
       60 Garden Street, Cambridge, MA 02138 USA 
\email{mmachacek@cfa.harvard.edu, cjf@head-cfa.cfa.harvard.edu,
      wrf@head-cfa.cfa.harvard.edu}}
\begin{abstract}
We present results from a $25$~ksec CHANDRA ACIS-S observation of
galaxies NGC 4438 and NGC 4435 in the Virgo Cluster. X-ray emission in
NGC 4438 is observed in a $\sim 700$\,pc nuclear region, a $\sim 2.3$\,kpc
spherical bulge, and a network of filaments extending $4-10$\,kpc to 
the west and southwest of the galaxy. 
The X-ray emission in all three regions is
highly correlated to similar features observed in H$\alpha$.
Spectra of the filaments and bulge are well represented by a $0.4$\,keV
MEKAL model with combined $0.3-2$\,keV intrinsic luminosity  
$L_{\rm X} = 1.24 \times 10^{40}$\ergs, electron densities 
$\sim 0.02-0.04$\cmc, cooling times of $400 - 700$\,Myr and X-ray 
gas mass $\lesssim 3.7 \times 10^8\Ms$. In the nuclear region
of NGC 4438 X-ray emission is seen from the nucleus and from two 
outflow bubbles extending $360$\,pc ($730$\,pc) to the 
northwest (southeast) of the 
nucleus. The spectrum of the northwest outflow bubble plus nucleus is well
fitted by an absorbed ($n_H = 1.9^{+1.0}_{-0.4} \times 10^{21}$\cms)
$0.58^{+0.04}_{-0.10}$\,keV MEKAL plasma model 
plus a heavily absorbed ($n_H = 2.9^{+3.1}_{-2.0} \times 10^{22}$\cms) 
$\Gamma = 2$, power law component. The  
electron density, cooling time, and X-ray gas mass in the 
northwest outflow are
$\sim 0.5$\cmc, $30$\,Myr and $3.5 \times 10^6\Ms$. 
Weak X-ray emission is observed in the central region of NGC 4435 with
the peak of the hard emission coincident with the galaxy's optical
center; while the peak of the soft X-ray emission is displaced 
$316$~pc to the northeast. The spectrum of NGC 4435 is well fitted by a 
non-thermal power law plus a thermal component from $0.2-0.3$\,keV
diffuse interstellar medium gas. 
We argue that the X-ray properties of gas outside the nuclear region
in NGC 4438 and in NGC 4435 favor a high velocity, off-center
collision between these galaxies $\sim 100$\,Myr ago; 
while the nuclear X-ray emitting outflow gas in NGC 4438 
has been heated only recently (within $\sim 1-2$\,Myr) by 
shocks ($v_s \sim 600$\kms) possibly powered by a central 
active galactic nucleus.
\end{abstract}



\section{Introduction}
\label{sec:introduction}

It is well established that the stellar populations in galaxies in the
cores of rich clusters have undergone substantial changes on
relatively recent timescales ($z \lesssim 0.5$). Butcher \& Oemler 
(1978, 1984) showed that clusters of galaxies at high redshift
($z>0.2$) generally had a larger fraction of blue galaxies compared to
that observed in nearby clusters. While morphologically these galaxies
appeared to fall on the normal Hubble sequence, spectroscopically many
of them were quite different from those observed in the
present epoch. Dressler \& Gunn (1983) found that
these blue galaxies belonged spectroscopically to three general
classes: galaxies with ongoing star formation, galaxies containing
active galactic nuclei (AGN) with
characteristic broad and/or high-excitation line spectra, and E+A
galaxies with strong Balmer absorption and little emission,
characteristic of an early type galaxy in the post-starburst phase. Using 
the Hubble Space Telescope (HST), Dressler \etal (1994) and Couch \etal
(1998) found that most of the blue galaxies in distant clusters were
of the first type, spirals or irregulars undergoing active star
formation, and, as redshift decreased, these galaxies were found to 
increasingly avoid cluster cores. While the fate of these blue
spiral and irregular galaxies is still debated, i.e. whether they
simply fade from view or transform their morphological type, it is
widely thought that their disappearance within rich cluster cores by the 
present epoch is driven by changes in their star formation rates and
that the causes responsible for the observed evolution are
more likely due to galaxy interactions with the cluster environment, 
predominantly within the dense cluster core, than to factors intrinsic
to the galaxies themselves (e.g. Poggianti \etal 1999 and references
therein).

A number of mechanisms for the interactions of galaxies within the
rich cluster environment have been suggested. Morphological evidence
from high resolution HST observations (Dressler \etal 1994; Couch
\etal 1998) show that $20\%$ of cluster galaxies, particularly those
very blue galaxies undergoing or having just recently completed a
starburst, show signatures of major mergers, i.e. interactions
between galaxies of similar size, confirming the 
suggestion by Lavery \& Henry (1988) that in some cases
mergers play an important role
in triggering star formation and changes of morphological
type. However, major mergers also are expected 
to completely disrupt the disk structure of the merger participants
and yet disk structures persist, albeit many show evidence of structural
distortion. For these galaxies some combination of minor mergers,
galaxy-galaxy tidal interactions (Moore 1996), tidal interactions
between the galaxy and the cluster potential (Byrd \& Valtonen 1990)
and/or gas depletion mechanisms such as ram pressure stripping (Gunn \& Gott
1972; Bekki \etal 2002) and turbulent or 
viscous stripping (Nulsen 1982; Quilis \etal 2000)
of the galaxy's interstellar medium (ISM) due to the motion of the
galaxy through the hot, intracluster medium 
(ICM) are likely to be important. Furthermore, 
galaxies, in dynamically young clusters far from virialization, are more
likely to show enhanced star formation or AGN activity than those in
relaxed, virialized clusters (Couch \etal 1998; Ellingson \etal 2001;
Miller 2003). This  suggests that dynamical changes in the cluster
potential or in the properties of the ICM due to subcluster mergers
significantly affect the evolution of the galaxies moving through
them.  What is clear from the data is that no single mechanism is able
to explain the observed transformation of galaxy colors and types in 
clusters. Rather, several processes may  act in concert to deplete the
gas content of galaxies, regulate their star formation, and drive
their evolution. 

Studies at a variety of wavelengths are essential in
order to determine the relative contributions of the different 
physical processes acting on galaxies within clusters. Simulations
show that different physical processes produce quite different density
and temperature structures within the gas (Abadi \etal 1999; Quilis
\etal 2000; Schulz \& Struck 2001; Vollmer \etal 2001). Thus high resolution
X-ray observations may help distinguish between competing
scenarios. While it is not yet possible to measure detailed X-ray
characteristics within blue galaxies in distant clusters, there are
spiral and irregular galaxies in present-epoch clusters that appear to
be analogs to the high redshift blue galaxies. In this paper we
present high resolution X-ray results from a $25$~ks observation of 
one such galaxy, NGC 4438 in the Virgo Cluster, taken with the Chandra
X-ray Observatory. 

The Virgo Cluster is a particularly good laboratory in which to study
the effects of cluster environment on galaxies. It is a fairly rich, 
nearby ($D=16$\,Mpc) cluster with a significant ICM and 
population of infalling spirals. It also is
dynamically young, showing signs of imminent or ongoing merging between
a small $\sim 1-3 \times 10^{13}\Ms$ group associated with the
elliptical galaxy M86 ($v=-244$\kms) falling from behind into the
major ($\sim 2\times10^{14}\Ms$) cluster component associated
with the dominant elliptical galaxy M87 ($v=1307$\kms) located at the
cluster core (e.g. Schindler \etal 1999). NCG 4438 
($12^h27^m 45.6^s$, $+13^\circ 00' 32''$, J2000) is a large SA/0 peculiar
spiral galaxy located in projection only $58'$ ($\sim 270$~kpc) 
to the northwest of M87  in the direction of 
M86. Given its extremely low line of sight velocity ($v=71$\kms), NGC
4438 may well be associated with the M86 subcluster. The nearest
galaxy (in projection) to NGC 4438 is NGC 4435, a compact SB galaxy
with line of sight velocity $v=801$\kms located $4'.3$ to the north
and west. 

NGC 4438 is the most environmentally ``damaged'' galaxy in
the Virgo cluster.
Its highly disturbed optical morphology shows stellar tidal debris
displaced well out of the disk in the
direction of NGC 4435. This general westward asymmetry is also found
in all tracers of the interstellar gas. Chincarini \& de
Souza (1985) observed a westward extension in H$\alpha+$[N II] emission. 
Higher resolution measurements (Keel \& Wehrle 1993, Kenney \etal
1995) found that this extension formed filamentary fingers of 
H$\alpha+$[N II] emission extending westward from the disk. 
Molecular gas identified by CO emission was found displaced $68''$ to the
west (Combes \etal 1988), coincident with a bright region of H$\alpha$
emission, as well as in the galactic center (Combes \etal 1988; Kenney \etal
1995). X-ray observations using 
the Einstein satellite (Kotanyi \etal 1983; Fabbiano \etal 1992)
showed both compact emission in the center of the
galaxy and extended emission to the west, as do $1.4$~GHz radio
continuum maps by Hummel \etal (1983). What little cold  HI gas that
is detected lies in two major clumps west of the galactic nucleus, one 
$70''$ to the west (Cayatte \etal 1990) between two H$\alpha$
filaments in a region of low X-ray surface brightness 
(Kenney \etal 1995), and the other about halfway
between NGC 4438 and NGC 4435 (Hibbard \& van Gorkom 1990). Together
these observations indicate highly disturbed interstellar gas
characterized by a wide range of density and temperature. 

Several possible interaction scenarios have been proposed to explain
the asymmetric, extended features seen in NGC 4438. One possibility
is that the violent disruption of the gas
could be caused by ICM-ISM interactions, such as ram pressure
stripping, as NGC 4438 passes with high relative
velocity ($\sim 1200$\kms ) through the densest part of the ICM near 
the Virgo cluster core (Kotanyi \etal 1983; Chincarini \& de Souza 1985; 
Keel \& Wehrle 1993). 
Alternatively, Combes \etal (1988) suggested that the pattern of
stellar tidal debris was more consistent with a high velocity
collision between NGC 4438 and NGC 4435; while the western, asymmetric
gas features could have been caused by ISM-ISM interactions in such a 
collision (Kenney \etal 1995).

The nuclear region in NGC 4438 is equally intriguing. On the
basis of  low  H$\alpha/[{\rm N II}]$ ratios, indicating that 
photoionization by hot stars is not the dominant source of ionization,
and the observation of a weak broad component in the H$\alpha$
emission line profile, denoting the presence of high
velocity gas (Ho \etal 1997c), NGC 4438 was classified as having a  
LINER$1.9$ or dwarf Seyfert nucleus. However, if photoionization by an 
AGN dominates, the X-ray emission should be strongly
correlated with the H$\alpha$ emission in both the hard and soft
energy bands (Ward \etal 1988; Koratkar \etal 1995). Although the
nuclear region is a strong source of soft, extended 
X-ray emission, as observed with Einstein ( Kotanyi \etal 1983), ASCA
(Terashima \etal 2000), and ROSAT HRI (Roberts \& Warwick
2000; Halderson \etal 2001), the ratio of the 
$2-10$~keV X-ray luminosity to the H$\alpha$
luminosity was found to be  anomalously low for its  LINER classification 
(Terashima \etal 2000). Using both broadband  and narrowband HST images of
the central $18''$ of the galaxy, Kenney \& Yale (2002) found bipolar
shells indicative of gas outflows straddling a compact ($1''$) nuclear
source. They observed that the low H$\alpha/[{\rm N II}]$ ratios 
originated predominantly from the edges of the shells from shock
heated gas produced by the outflows interacting with the surrounding 
ISM rather than from the nuclear source. A similarly complex
morphology was  found in 
$1.49$ and $4.86$~GHz radio continuum maps (Hummel \& Saikia 1991) with
elongated structures in general agreement with the optical shells and
a weak radio source at the nucleus. The nature of the nuclear source
that powers the observed gas outflows remained unclear. 
Some properties of the outflows were found consistent with those 
expected from a very compact nuclear starburst, while others were
better characterized as originating from a true AGN.

In this paper we reexamine these issues using Chandra observations of 
NGC 4438 and NGC 4435. We study the interaction of NGC 4438 with 
its environment through X-ray measurements of the properties of the 
disturbed, extended gas. We also probe with high spatial resolution 
the X-ray properties of the galactic center to clarify the nature of 
the nuclear source. This paper is organized in the following way.  
In Section \ref{sec:obs} we detail the 
observations and the data reduction and processing procedures.
In Section \ref{sec:results} we discuss our main results. First 
we exploit the spatial resolution of Chandra to compare the X-ray to 
optical morphology of the field containing both
NGC 4438 and its companion galaxy NGC 4435 as a whole. Then in Section 
\ref{sec:outer} we compare the X-ray properties of gas in the extended
filamentary features westward of NGC 4438 to those of gas in the bulge. 
We also compare our X-ray results to other wavelength
bands. We show that the X-ray emission in the extended features, as well
as the bulge, is well characterized by a $0.4$\,keV plasma model.  The 
extended X-ray features found asymmetrically out of the plane of the disk 
to the west and southwest are 
highly correlated with the H$\alpha+$[N II] emission filaments, 
supporting the 
ISM-ISM interaction hypothesis of Kenney \etal (1995) for their
origin. In Section \ref{sec:core} we study the nuclear emission region
consisting of two outflow shells and a compact nuclear
source. We show that X-ray emission from the gas in the outflow is 
consistent with shock heating from a relatively recent outburst from
the nucleus and that the data favor a highly obscured AGN over a
compact starburst for the nuclear source. 
In Section \ref{sec:n4435}
we investigate the X-ray emission from NGC 4438's companion galaxy 
NGC 4435. We find evidence in NGC 4435 for a power law nuclear source 
as well as a weak, cool
extended thermal component.
We briefly discuss the point source population of the field 
in Section \ref{sec:points} and
summarize our results in Section \ref{sec:conclude}. Unless stated
otherwise, all fluxes are observed fluxes, while all luminosities are
intrinsic, i.e. have been corrected for absorption. All errors are 
$90\%$ confidence limits and coordinates are J2000. 
Taking the distance to NGC 4438 as that of the Virgo Cluster core
($16.1$\,Mpc, Kelson \etal 2000), $1''$ 
corresponds to a distance scale of $77$~pc. 
  

\section{Observations and Analysis}
\label{sec:obs}

Our data consist of a $25$~ksec exposure of NGC 4438 taken by the 
Chandra X-ray Observatory on 29 January 2002. The data were obtained
using the Advanced CCD Imaging Spectrometer array
(ACIS, Garmire \etal 1992; Bautz \etal 1998) operating in very faint
mode to improve background rejection. The  focal
plane temperature of the instrument was $-120^\circ$~C throughout the 
observation. Our analysis of NGC 4438 is based mainly on data near the 
aim point from the back-illuminated CCD S3 chip, although the imaging
and spectral analysis of nearby NGC 4435 utilized data from the S2
chip as well. Each CCD chip is a $1024 \times 1024$ pixel array where each
pixel subtends $0''.492 \times 0''.492$ on the sky. 

The data were analyzed using the standard X-ray processing packages, 
CIAO $2.3$,  FTOOLS and XSPEC $10.0$. Cosmic ray
contamination was minimized by filtering out events with bad
grades ($1$, $5$, and $7$) and those tagged as bad by the VFAINT data 
acquisition mode due to significant flux in border pixels of the $5
\times 5$ event islands. The use of VFAINT mode improves background 
rejection by as much as a factor of $3$ in the back illuminated S3
chip for the soft X-ray energies ($\sim 0.3$~keV) important to this 
observation. Bad pixels and columns also were removed in the standard 
manner. The light curve was checked for flares, i.e. periods of
anomalously high background, and none were found. Since the spectrum
of NGC 4438 is known to be soft, special care was 
taken to model detector response at low energies. In 
addition to the standard modeling of instrument response functions for
spectral analysis, we included corrections for the declining
efficiency of the ACIS detector, significant for energies below
$1$~keV, due to the buildup of contaminants on the optical filter and
explicitly checked our spectral fits for sensitivity to this
correction.

Blank sky background sets appropriate for the date of 
observation were used in the imaging analyses 
(see http://cxc.harvard.edu/ciao/threads/acisbackground/).  
We applied the same cleaning
algorithms to the data that were applied to the background sets. This
resulted in the additional removal
of short intervals 
at the beginning 
and end of the observation, reducing the useful exposure time to 
$24,891$~s.  Identical spatial and energy filters were applied to
source and background data throughout,
so that the background normalization is set by the ratio of exposure
times. We checked this normalization by comparing count rates
in the high energy ($9$-$14$~keV) band, where particle background 
is expected to dominate, to that predicted by the normalized blank field
background sets and found that they differ 
by $ \lesssim 6\%$. We adopt this difference as our relative
uncertainty in the background level for this method. 

Blank sky background subtraction is, in most cases, the preferred
background subtraction technique for ACIS-S observations 
of extended sources since it
models correctly the variation of the background across the 
S3 chip. In the spectral analysis of the extended emission from NGC
4438, however, it is important to account for the additional
background due to the presence of the hot Virgo intracluster gas.
We separated the cluster ICM component from the non-nuclear extended 
galactic ISM emission in two ways.  First (Method B) we used blank sky 
background subtraction and modeled the intracluster gas explicitly in our
spectral fits by including a fixed MEKAL thermal plasma model
component appropriate for cluster gas at the location of NGC 4438, 
i.e $58'$ ($272$~kpc) from M87.
We used our  current data from a $95''$ circular region centered at 
 ($12^h27^m 35.4^s$, $+12^\circ 57' 11''$) on chip S3  away from the 
NGC 4438 emission and blank sky background subtraction to fix the 
cluster parameters. We found the cluster data well represented by a 
MEKAL plasma model with $kT = 2.1 \pm 0.3$\,keV 
($\chi^2/{\rm dof}=80.6/77$) for fixed 
abundance $0.3\,\Zs$ (Ohashi \etal 1998) 
and Galactic absorption ($2.64 \times 10^{20}$\cms). The normalization
of the cluster component in the spectral analyses of extended emission
from NGC 4438 was fixed by the ratio of the areas of the  
galaxy to cluster extraction regions. We also used our model of
the cluster gas to determine the cluster electron density in the
vicinity of NGC 4438. Assuming a spherical $\beta$ model for the gas 
density distribution of the Virgo cluster with $\beta=0.47$ and core 
radius $2'.7$, we found an electron density for the hot ICM local to 
NGC 4438 of $\sim 5.7 \times 10^{-4}$\cmc. 
These  Virgo cluster model parameters and the inferred cluster electron
density are in general agreement with previous measurements 
(Ohashi \etal 1998, Schindler \etal 1999).
Second (Method L) we took
the same region used to fit the cluster gas as a local background region 
to directly subtract the background plus cluster emission. We examined
blank sky background sets to verify that variations
in the background observed between the source region
and local background region were small, $\lesssim 3\%$ for the 
$0.3-2$\,keV energy range used in these spectral fits. 
We used the local background region to subtract both the standard 
background and Virgo cluster contribution and performed  spectral fits 
to the outer extended emission from NGC 4438. 

For the nuclear emission region, additional backgrounds consist of
both the Virgo cluster gas and the outer extended emission from NGC 4438 
itself. For this smaller extraction region, background variation 
across the S3 chip is not a concern. We used a local background 
annulus of $7''.4$ inner and $20''.7$ outer
radii centered on the $5''$ nuclear region 
to separate the nuclear emission 
from the cluster and outer galactic components.
	
\section{Results}
\label{sec:results}

\begin{figure}[t]
\plotone{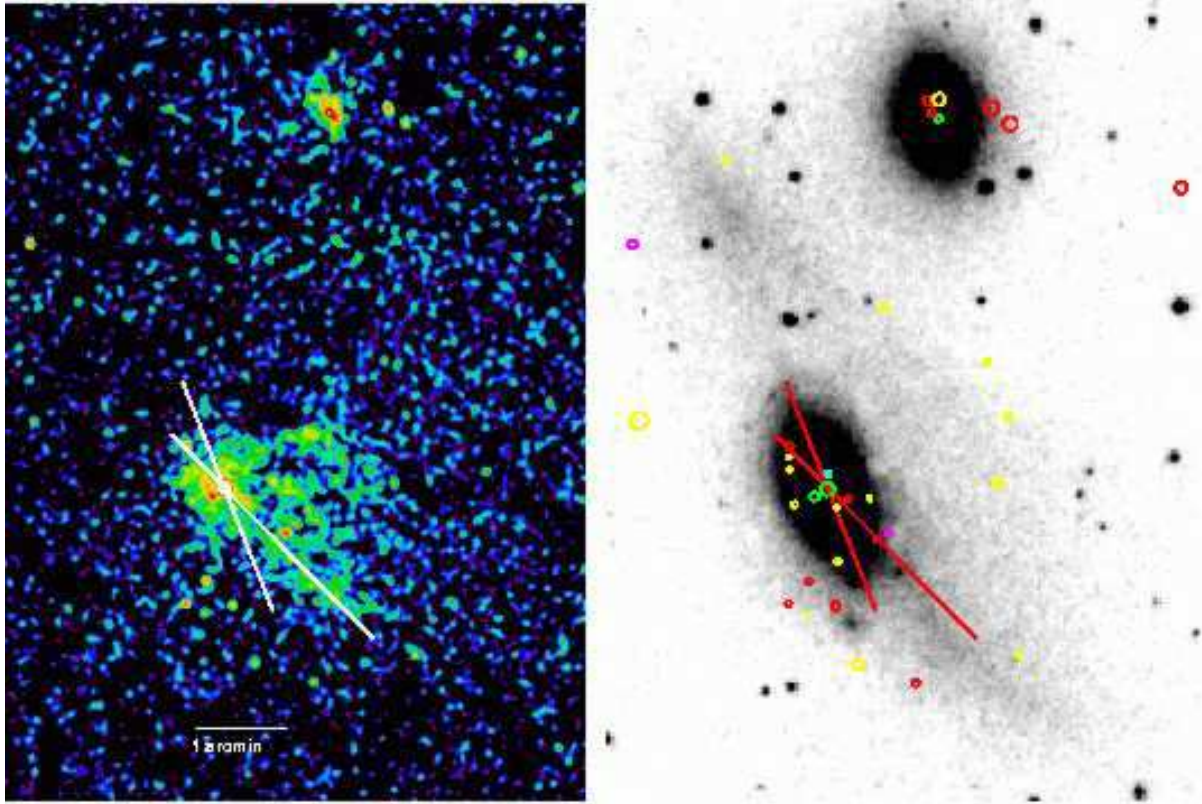}
\caption{(left) Chandra 
$0.3-2$~keV band image of the $6'.6 \times 9'.4$ field containing 
galaxies NGC 4438 and NGC 4435. North is up and East is to the left.
The X-ray image has been Gaussian smoothed ($\sigma=2''$), background
subtracted and exposure corrected. The image stretch is logarithmic
from $4 \times 10^{-9}$ to $4 \times 10^{-6}$\expunit.
(right) DSS optical ($\lambda6450$)
image of the same field taken with 
the Palomar $48$ inch Schmidt telescope and matched in WCS
coordinates with the Chandra image. The image stretch is linear from
$1000 - 5000$ counts\,arcsec$^{-2}$ and pixel size is $1''.7$.
Broadband ($0.3 - 10$\,keV) X-ray point sources are superposed as
circles on the optical image with colors denoting source strength 
in the hard ($2.0-10$\,kev) and soft ($0.3-2$\,keV) energy bands as 
follows: 
magenta denotes $\ge 10$ net counts in both the
hard and soft bands;
red (cyan) denotes $\ge 10$ net counts in the soft (hard) band but not 
both; yellow denotes $ < 10$ counts in either of the narrower bands. 
Green circles denote  nuclear outflow bubbles in NGC 4438 and the hard 
nucleus in NGC 4435. Crossed lines show the 
directions of the major axis of NGC 4438 and the leading X-ray
filament. The horizonal angular scale bar represents $1'$. 
}
\label{fig:bigmap}
\end{figure}

In Figure \ref{fig:bigmap}  we compare a Palomar DSS optical
($\lambda6450$) image of a $6'.6 \times 9'.4$ region containing NGC
4438 and NGC 4435 (right panel) to the $0.3-2$~keV Chandra X-ray 
image (left panel) of the same region matched in WCS coordinates. 
The Chandra data have been smoothed with a
$\sigma=2''$ Gaussian and background subtracted.
Telescope vignetting  and spatial efficiency variations  have
been corrected by means of an exposure map generated using standard  
CIAO tools assuming a fixed energy of $0.85$ keV corresponding
approximately to the peak of the X-ray spectrum.

The highly disturbed morphology of NGC 4438 is apparent in both the
optical and X-ray images in Figure \ref{fig:bigmap}. 
The optical image shows tidally disrupted stellar
features extending $\sim 4'.4$ to the northeast and $\sim 3'.6$ to the 
southwest of the galactic center. An additional tidally distorted
stellar fold makes an angle of $\sim 27^\circ$ with respect to the
northeastern feature in the direction of NGC 4435; while the southwestern
structure forms a nearly linear stellar ridge. The X-ray emitting
gas is equally disturbed. The X-ray emission can be separated into
three major components: a strongly emitting nuclear 
region, a roughly spherical (radius $\sim 2.3$\,kpc) bulge emission
region surrounding the nucleus, and a network of 
filamentary features $\sim 1$\,kpc thick extending $\sim 4-10$\,kpc to the
west, well out of the galactic disk. The leading (most southern) X-ray 
filament lies $\lesssim 8''.7$ west of the southwestern optical
stellar tidal ridge and is oriented  
$\sim  23 ^\circ$ counterclockwise from the major axis of the galaxy
as shown by crossed lines in both panels of Figure \ref{fig:bigmap}.
We discuss the detailed properties of these diffuse filamentary
structures in Section \ref{sec:outer}.
The positions of two extended regions of X-ray emission located near 
the galactic center are superposed as green circles on the optical image. 
 We will show in Section \ref{sec:core} that these regions correspond
to the nuclear outflow regions observed by Kenney \& Yale (2002) in 
H$\alpha$+[N II] emission.
NGC 4435 also shows asymmetric emission about its center that will 
be discussed further in Section \ref{sec:n4435}. The locations of 
broadband ($0.3-10$\,keV)
point sources identified by a multiscale wavelet
decomposition algorithm are shown as circles in Figure
\ref{fig:bigmap}, color coded to denote source strength. These will be 
discussed in Section \ref{sec:points}. 

\subsection{NGC 4438: The X-ray Filaments}
\label{sec:outer}

\begin{figure}[t]
\epsscale{1.0}
\plottwo{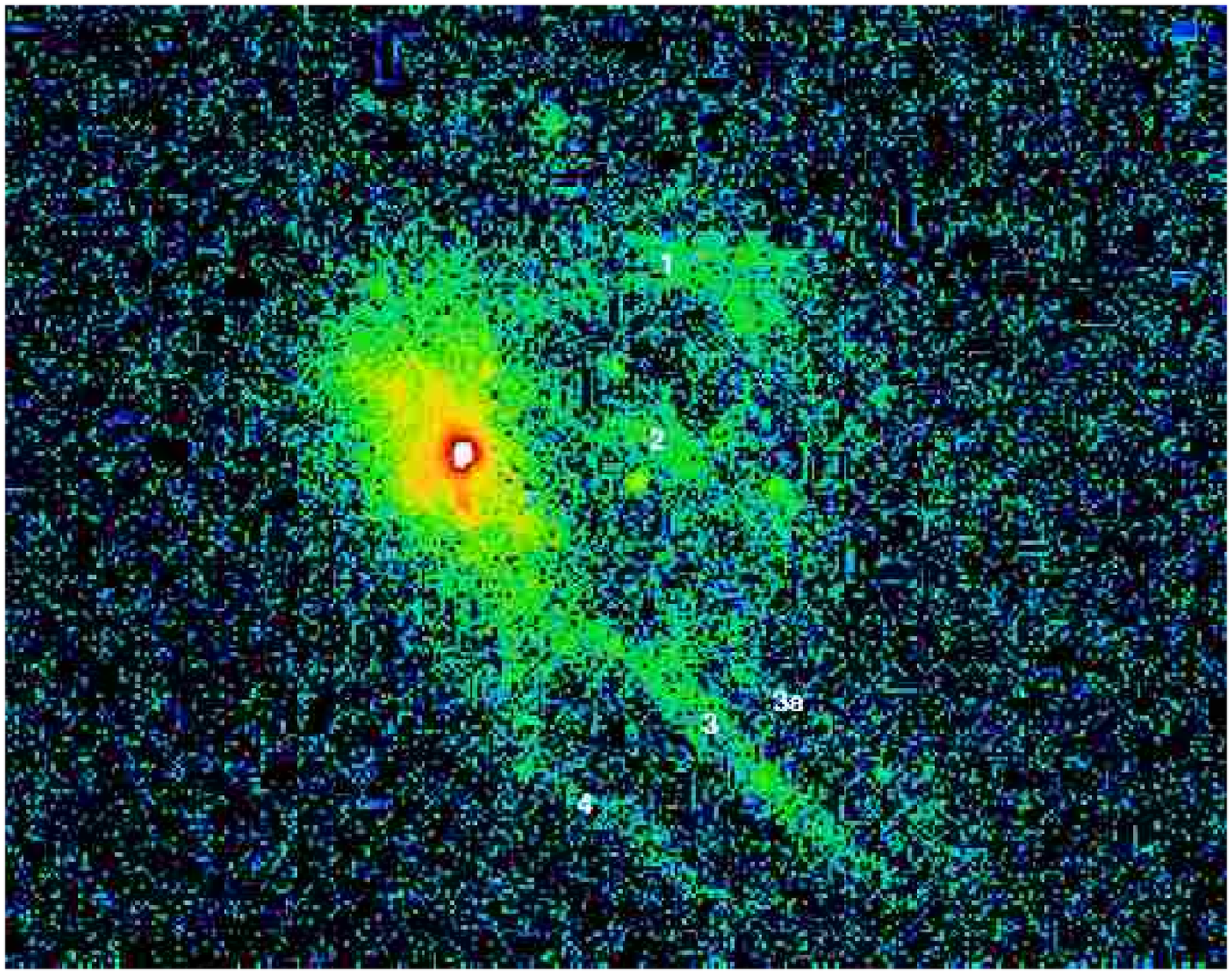}{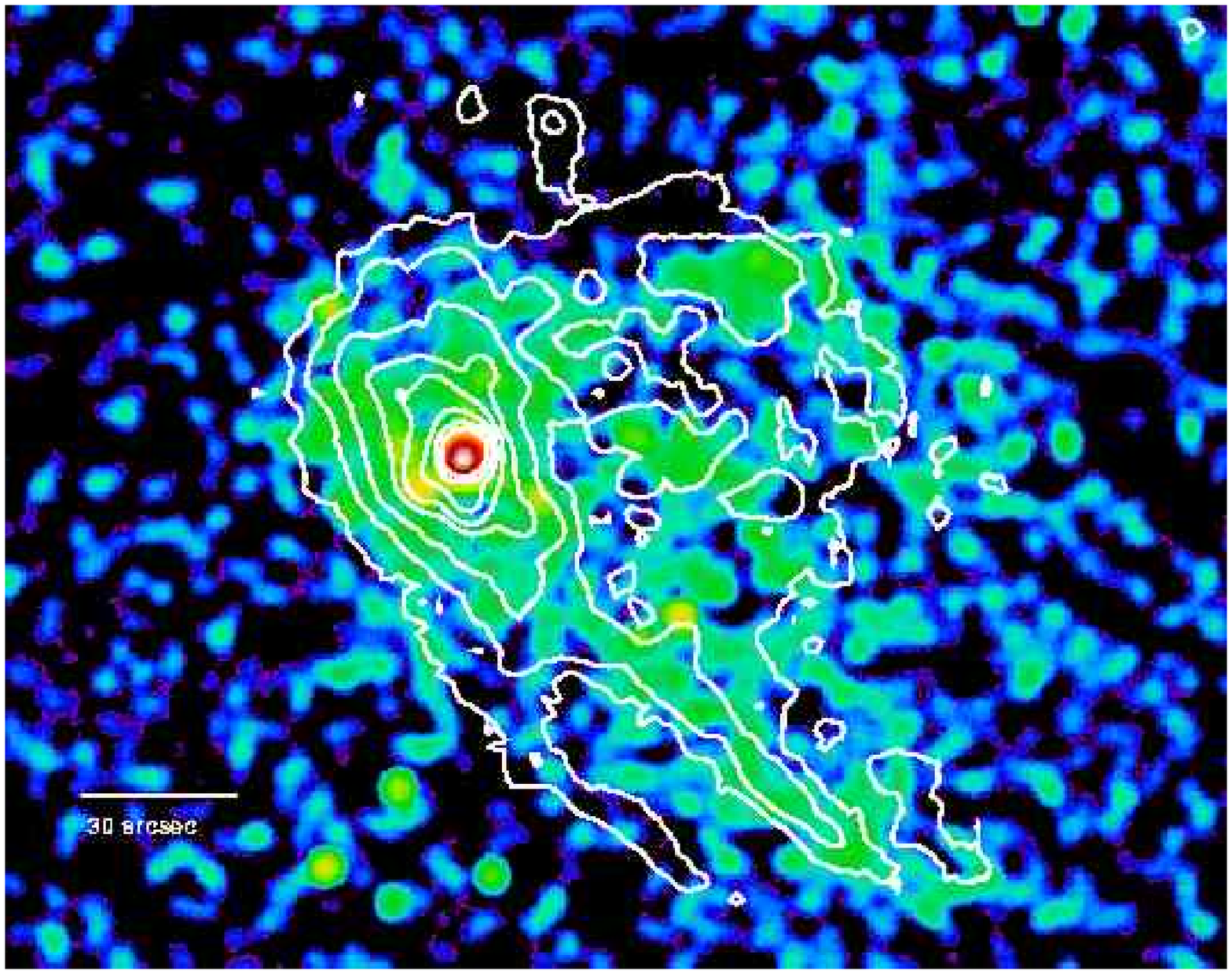}
\caption{(left) KPNO H$\alpha+$[N II] image of NGC 4438 (Kenney \etal
1995). The image stretch is logarithmic from $0.05\% -100\%$ of the
peak value. Contours superposed on the Chandra image to the right are
$0.0096\%$, $0.24\%$, $0.48\%$, $1.4\%$, $2.4\%$, $4.8\%$ of the peak
value.  (right) Chandra image of NGC 4438 in the $0.3-2$~keV energy
band with H$\alpha$ contours superposed. 
The Chandra image has been Gaussian smoothed ($\sigma=2''$), background
subtracted and exposure corrected. 
The image stretch is logarithmic
from $4 \times 10^{-9}$ to $4 \times 10^{-6}$\expunit.
The horizontal angular scale bar is $30''$ and in both images North is
up and East is to the left. 
}
\label{fig:Halpha}
\end{figure}

Diffuse X-ray emission from 
NGC 4438, as with other ISM tracers, is displaced to the west of the 
galaxy out of the plane of the disk.
In Figure \ref{fig:Halpha} we
compare the X-ray emission from NGC 4438 to the $0''.77$ resolution
H$\alpha+$[N II] map of Kenney \etal (1995) and find a striking 
correspondence between the X-ray and H$\alpha+$[N II] features. 
The H$\alpha$ filaments $1$, $2$ and $3$ (labeled
clockwise in the left panel of Figure \ref{fig:Halpha} from the 
northern-most filament as in Kenney \etal 1995), or equivalently 
the ``bubble'' (composed of filaments $1$
and $2$) and linear filament (filament $3$) from 
Keel \& Wehrle (1993), coincide with similarly shaped X-ray
features. Although we detect faint X-ray emission from a region
coincident with the faint H$\alpha$ filament $4$ identified by 
Kenney \etal (1995), it appears patchy rather than forming a 
linear structure as seen in H$\alpha$.  
We do see faint X-ray emission in the form of a short, linear 
extension (labeled filament $3a$) similar to a weak H$\alpha$ feature 
described by Kenney \etal (1995)
roughly parallel to filament $3$ beginning
$\sim 13''$ to the northwest of that filament near a bright
X-ray point source.  We find, in agreement with previous Einstein HRI
observations (Kotanyi \etal 1983), that the brightest region of
X-ray emission in filament 1, located $\sim 70''$
to the northwest of the nucleus, overlaps both the region of
weak extended CO emission observed by Combes \etal (1988) and the  
dominant H$\alpha+$[N II] emission in this filament (Kenney \etal 1995). 

\subsubsection{Mean Spectral Properties}
\label{sec:mean}

In order to obtain an average spectrum for the diffuse gas,
we first fit the extended 
emission in the mean by choosing a circular 
extraction region $95''$ in radius centered at 
( $12^h 27^m 42.3^s, +13^{\circ} 00' 13''$) containing 
the bulge and filamentary features,
but with point sources and two circular regions 
of radius $4''.4$ and $3''$  
containing the northwestern and southeastern nuclear outflow regions
removed. Both blank sky and local background subtraction methods 
were used to check that our results for the emission from NGC 4438 
were not dependent on our
treatment of the Virgo cluster emission.  
Spectra were analyzed using XSPEC $10.0$ with the data grouped to 
require a minimum of $20$ counts per spectral bin. 
The results of these fits are summarized in Table \ref{tab:spectra}.

\begin{deluxetable}{ccccc}
\tablewidth{0pc}
\tablecaption{Mean spectral fits to the diffuse gas in NGC~4438\label{tab:spectra}}
\tablehead{
\colhead{Method/Model}  &\colhead{B/MEKAL} &\colhead{B/MEKAL} &
\colhead{L/MEKAL} & \colhead{L/MEKAL}} 
\startdata
source counts &$3390$ &$3390$ &$2238$ & $2238$  \\
energy (keV) & $0.3-3$ & $0.3-3$ & $0.3-2$ & $0.3-2$  \\
$kT$ (kev)   & $0.37^{+0.03}_{-0.02}$ & $0.38^{+0.03}_{-0.02}$
&$0.42^{+0.07}_{-0.04}$ & $0.41^{+0.10}_{-0.06}$  \\
$A$ ($\Zs$)  & $0.14^{+0.05}_{-0.03}$ & $0.18^{+0.13}_{-0.07}$
&$0.11^{+0.05}_{-0.03}$ & $0.11^{+0.05}_{-0.04}$\\
N$_{\rm H}$($10^{20}$\cms) &$2.64^f$ &$\leq 3.6$ &$2.64^f$ & $\leq 9.1$ \\
$\chi^2/{\rm dof}$   &$140/94$ &$139/93$ &$103/82$ & $103/81$ \\
\enddata
\tablecomments{Method B (blank sky background subtraction) fits both cluster
and galaxy components. Method L (local background subtraction)  
fits the galaxy component alone. Superscript $f$ denotes a fixed parameter.
}
\end{deluxetable}
When deriving
the gas temperature
and abundance for the diffuse emission using  blank sky background
data (Method B), we modeled the 
contribution of the hot Virgo cluster gas explicitly by including a 
MEKAL plasma emission model component with 
parameters fixed by our fit to the Virgo cluster gas 
at the radial distance of NGC 4438. A second 
MEKAL component was used to characterize the galactic emission and
a common hydrogen absorbing column was assumed for both components.
The absorbing column was initially fixed at the Galactic value 
($2.64 \times 10^{20}$\cms). The temperature, metallicity, and 
normalization of the second MEKAL component were fit (Model MEKAL). 
We restricted the energy range to 
$0.3-3$\,keV where there were significant source counts compared to
the background.
A temperature $kT=0.37^{+0.03}_{-0.02}$\,keV and 
abundance $A=0.14^{+0.05}_{-0.03}\,\Zs$ ($90\%$ CL) gave an acceptable
fit to the mean spectrum of the diffuse X-ray emitting gas. 
We checked the stability of this fit by allowing  the absorbing
column to vary and show the results in Table \ref{tab:spectra} column
$3$. The temperature of the gas 
remains well determined ($kT=0.38^{+0.03}_{-0.02}$\,keV), 
although the absorbing column and abundances are less well constrained 
(N$_{\rm H} \leq 3.6 \times 10^{20}$\cms,
  $A=0.18^{+0.13}_{-0.07}\,\Zs$). 
The fits agree within their $90\%$ confidence limits.

The total $0.3-2$\,keV flux for this region is
$4.93 \times 10^{-13}$\ergscm with NGC 4438 contributing
$68\%$ ($3.35 \times 10^{-13}$\ergscm) and the cluster 
accounting for the remaining $32\%$ ($1.58 \times 10^{-13}$\ergscm) in
this energy band. After correcting for Galactic absorption, the $0.3-2$\,keV 
intrinsic luminosity 
of the diffuse emission at a distance of $16.1$\,Mpc is 
L$^{total}_{\rm X} = 1.81 \times 10^{40}$\ergs with 
$1.25 \times 10^{40}$\ergs 
from NGC 4438
and 
$5.6 \times 10^{39}$\ergs from the ICM
component. 
Since the diffuse emission from NGC 4438 is soft, with 
only $\sim 1\%$ 
of its flux above $2$\,keV, 
the $2-10$\,keV luminosity is dominated by the 
hot cluster component. 
The diffuse gas in NGC 4438  contributes only  
$\sim 2\%$ ($7.1 \times 10^{37}$\ergs) of the total  
$2.9 \times 10^{39}$\ergs hard band emission. 

When  emission from the Virgo cluster ICM is
subtracted by selecting a local background region 
(Method L), the number of net ``source''
counts is reduced from $3390$ to $2238$ due to the
elimination of the Virgo cluster emission.
Since the diffuse regions of NGC 4438 contribute little
to the X-ray emission above $2$\,keV, we restrict 
the energy band for these spectral fits 
to $0.3-2$\,keV 
where the galaxy count rate is significantly above background. 
Holding the absorbing column fixed at the Galactic value, we
find the spectrum for the mean outer diffuse gas from 
NGC 4438 is well
fit by a single temperature MEKAL model with gas temperature 
$kT=0.42^{+0.07}_{-0.04}$\,keV and abundance $A=0.11^{+0.05}_{-0.03}\,\Zs$. 
This gives an X-ray flux in the 
$0.3-2$\,keV energy band for the diffuse gas from NGC 4438 of 
$3.35 \times 10^{-13}$\ergscm corresponding to an intrinsic X-ray luminosity 
in this band of $1.24 \times 10^{40}$\ergs, in good agreement with the 
results of Method B. Again, as shown in column $5$ of Table
\ref{tab:spectra}, the fit is stable if the column density is 
also allowed to vary.
 
We also tried fitting the data, using both Method B and Method L, with
a variable MEKAL model where we divided the elements into two groups,
the alpha elements (O, Mg, Si, Ca) and all others, and tied the element 
abundances in each group together in the fit. We then fit
the temperature and each group of abundances, while keeping the
absorbing column fixed at the Galactic value. Within their $90\%$ 
confidence limits, there was no significant difference in the
abundance for alpha and non-alpha elements. Similarly the temperature and 
abundances agreed within the $90\%$ confidence limits with those of the 
corresponding simple 
MEKAL model. We thus do not include them in Table \ref{tab:spectra}.

\subsubsection{Individual Features}
\label{sec:features}

\begin{figure}[t]
\epsscale{0.5}
\epsfig{file=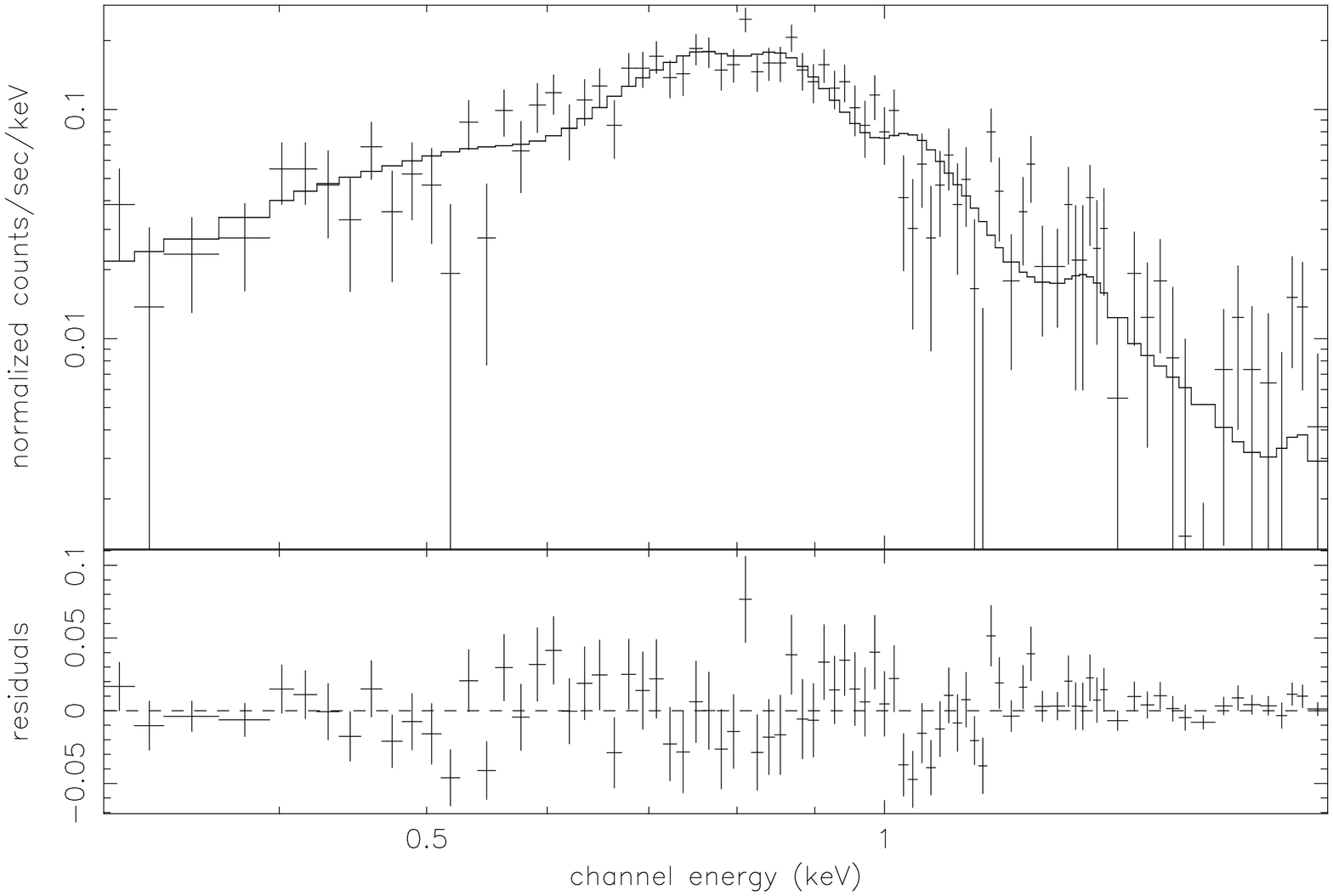,height=3in,width=3in,angle=270}
\hspace{0.3cm}\epsfig{file=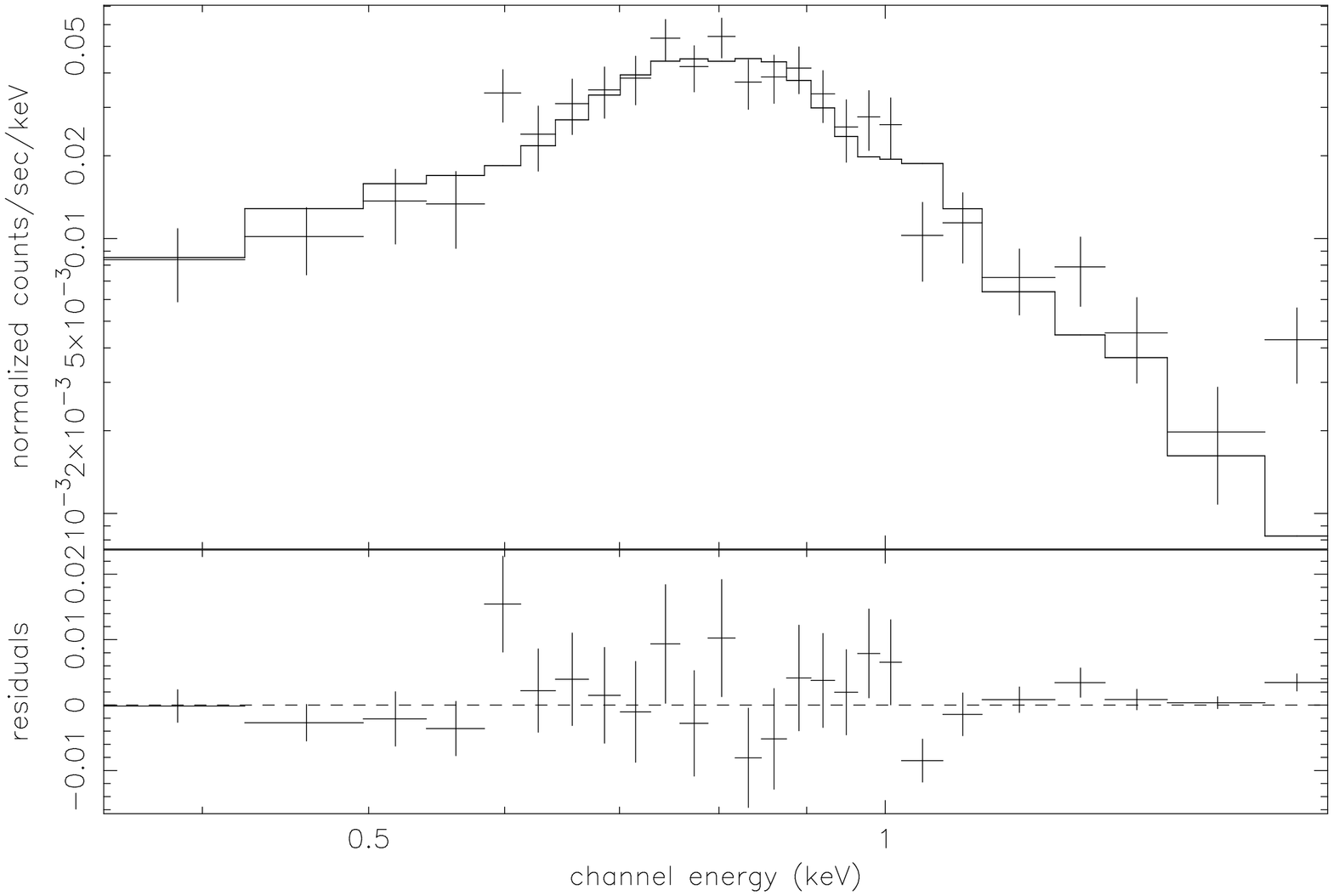,height=3in,width=3in,angle=270}
\caption{Spectra and best fit MEKAL models, using local
 background subtraction, for the extended diffuse
gas in NGC 4438 in the mean (left) and in the bulge (right).
}
\label{fig:diffusespec}
\end{figure}

To gain insight into the physical mechanisms responsible for 
the distorted morphology of NGC 4438, we compare the X-ray properties 
of the filaments to X-ray emitting gas in the bulge. 
We used a circular region of radius $30''$ ($2.3$\,kpc) about the 
nucleus, excluding point sources and the nuclear outflow regions,  
to extract the bulge emission and found $566 \pm 26$ source counts in
the $0.3-2$\,keV energy band (Method L), sufficient for an 
independent spectral fit. Using a
single temperature MEKAL model with fixed Galactic absorption, we found
a temperature $kT=0.42^{+0.09}_{-0.05}$\,keV and abundance 
$A=0.12^{+0.10}_{-0.04}\,\Zs$, in complete agreement with the mean fits
given in Table \ref{tab:spectra}. 
In Figure \ref{fig:diffusespec} we show spectra and the model fits 
for these two regions. 


Thus we found  no significant
difference in the temperature and abundances for gas in the bulge
compared to that in the mean ( and thus the filamentary extended 
features), even though the bulge gas contributes only 
$\sim 25 \%$ of the total extended diffuse emission. Thirty-four
percent  of the total extended emission comes from the
filaments/bubble, while the
remaining $41\%$ is from diffuse gas not associated with any feature.

In Table \ref{tab:features} we characterize the extended X-ray features 
by simple geometries, e.g a sphere for the bulge, cylinders 
(boxes in projection) for the
filaments and a spherical shell for the ``bubble''. The extent of each
feature is represented by the length scales for the assumed geometry, i.e.
the radius of a sphere, (radius $r$, length $l$) for cylinders, and 
(inner, outer) radii for the spherical shell. The cylinders are
assumed to be figures of rotation about their long axes, such that the 
line of sight extent of a filament is its diameter $2r$. Except for the
bulge, these individual features are too weak to independently
constrain all the X-ray spectral parameters. 
Instead we fixed the temperature and abundance at the 
best fit values given in Table \ref{tab:spectra} for 
the mean and allowed the 
normalizations to vary to determine the X-ray flux and luminosity of each
feature.  The spectral normalizations were then used to determine the 
emission measure for each region and, assuming gas uniformly fills 
the assumed geometry, to infer an rms electron density
$<n^2_e>^{1/2}$ and mass of X-ray emitting gas. 
The reader should note that if the gas does not uniformly fill the volume,
the electron density is increased by a factor $\eta^{-1/2}$ and the mass 
is decreased by a factor $\eta^{1/2}$ where $\eta \le 1$ is the filling 
factor for the region.
In Table \ref{tab:features} we list the fluxes, 
luminosities, average electron densities and X-ray emitting gas mass 
for the mean 
emission, bulge emission, and emission from each 
filament/bubble. 
Since the spectral fits in Table
\ref{tab:spectra} all agree within their $90 \%$ confidence limits, we
list in Table \ref{tab:features} only the results for the fixed MEKAL
model with local background subtraction. 
Errors on 
the inferred electron densities and hot gas masses are dominated by
uncertainties in the assumed geometries for each feature, rather than 
by uncertainties in the spectral fits.
\begin{deluxetable}{cccccccc}
\tabletypesize{\footnotesize}
\tablewidth{0pc}
\tablecaption{X-ray properties of extended emission in NGC 4438\label{tab:features}}
\tablehead{
\colhead{Region} & \colhead{Geometry}&\colhead{Size}&\colhead{NGC 4438} & \colhead{Flux} &
\colhead{L$^{ISM}_{\rm X}$} &\colhead{ $<n^2_e(gal)>^{1/2}$}&\colhead{M$_{gas}$} \\
 & &\colhead{(arcsec)} &\colhead{counts} &\colhead{($10^{-13} cgs$)} &\colhead{($10^{40}$\ergs)} & 
\colhead{($10^{-2}$cm$^{-3}$)}&\colhead{($10^7\Ms$)} }
\startdata
mean   &sphere  & $95$      &$2238 \pm 75$ & $3.35$ & $1.24$  & $0.8$
& $37$ \\
bulge & sphere & $30$ & $566 \pm 26$ & $0.83$ & $0.31$ & $2.3$ &$3.23$ \\
filament 1 &cylinder &($12,70$)  & $232 \pm 17$ &$0.37$  & $0.14$ &
$2.8$ & $1.15$ \\
filament 2 &cylinder &($7.6,54$) & $141 \pm 13$  &$0.22$ & $0.08$ &
$3.9$ & $0.50$  \\
bubble     & shell &($14,30$)  & $375 \pm 23$  &$0.57$  & $0.21$ &
$1.9$ & $2.58$  \\
filament 3 &cylinder &($7.4,120$)  & $269 \pm 19$  &$0.40$ & $0.15$
& $3.6$ & $0.98$  \\
filament 3a &cylinder &($4.6,77$) &  $55 \pm 9$ & $0.09$ & $0.03$  &
$3.4$  & $0.23$ \\
filament 4  &cylinder &($6,61$) &  $25 \pm 6$ & $0.06$ & $0.02$  &
$2.3$ & $0.22$ \\
\enddata
\tablecomments{Size is characterized by radius for spheres, 
(inner, outer) radii for the shell, and (radius, length) for
cylinders. 
The energy bandpass for flux and luminosity is 
$0.3-2$\,keV. Fluxes are 
observed. Luminosities are corrected for Galactic absorption.
``cgs'' denotes the flux unit \ergscm. 
} 
\end{deluxetable}

The rms electron densities for the X-ray 
emitting gas in the bulge and the filaments are found to be 
$\sim 0.02 - 0.04$\cmc, in agreement with estimates
by Kotanyi \etal (1983). The lower electron density ($0.008$\cmc) 
inferred from the diffuse emission 
as a whole (see the top row of Table \ref{tab:features}) is a
reflection of the fact that the X-ray emitting gas does not 
uniformly fill the volume, but much of it is distributed in the observed
filamentary or bubble-like features. 
If we assume a filling factor of
$1$ for the $95''$ spherical volume as a whole, we find a conservative
upper limit to the total extended X-ray emitting gas mass of 
$\lesssim 3.7 \times 10^8\Ms$, a factor of $2.4$ lower than the 
estimate by Kenney \etal (1995) using Einstein data and an approximate 
X-ray flux-to-mass conversion factor (Canizares \etal 1987).
On the other hand, if we assume that gas only resides in the bulge and
filamentary features, we derive a lower bound on
the total hot gas mass in NGC 4438 of $\gtrsim 6 - 7 \times 10^7\Ms$
(assuming uniform filling and the electron density derived for each
feature) with $3.2 \times 10^7\Ms$
 still bound to the bulge and, depending on the chosen geometry, 
$3 - 4 \times 10^7\Ms$ contained in the extended features.
Optical emission line spectra (Kenney \etal 1995; Chemin \etal 2003) 
show that the velocities of the cooler, optically emitting gas in
the extended filaments are within $200$\kms of the central velocity of 
NGC 4438. Assuming that the X-ray gas in each corresponding
feature shares a common flow velocity with that of the cooler gas, the gas
in these extended filaments may remain gravitationally bound to the
galaxy. 

Given the measured temperature ($\sim 0.4$\,keV), abundance 
($A \sim 0.1-0.2\,\Zs$) and inferred electron
density ($0.02-0.04$\cmc) for both X-ray emitting gas in the 
extended filamentary features and in the bulge, we can determine its 
characteristic cooling time. The cooling function for
this temperature and abundance is 
$\Lambda \sim 5 \times 10^{-24}$\,{\rm erg\,cm$^{3}$\,s$^{-1}$.
Thus, the cooling time for gas at these densities, 
\begin{equation}
t_{cool} = \frac{3 n kT}{2 n_e n_H \Lambda(T)}
\label{eq:cool}
\end{equation}
where $n$, $n_e$, and $n_H$ are the gas, electron,
and hydrogen number densities, $k$ is Boltzmann's constant and $T$ is
the temperature, is 
$ t_{cool} \sim 4 - 7 \times 10^8$\,yr.

It is also interesting to note that, using our derived values for the 
electron densities in the cluster and the bulge, the thermal pressure 
of the X-ray emitting gas in the bulge is only a factor $\sim 1.5$ 
larger than the ram pressure on NGC 4438 due to 
its motion through the Virgo ICM. The contribution of the thermal 
pressure of the
ICM is small compared to the contribution of the ram pressure at these 
velocities. Given the large uncertainties in the electron 
densities, this is consistent with local pressure equilibrium between
the leading hot bulge gas and the ICM wind. We can also compare the 
thermal pressure between the cooler gas in the optical filaments 
and X-ray gas. Kenney \etal (1995) and 
Keel \& Wehrle (1993) infer electron densities 
in the optical filaments of 
$\sim 500$\cmc and $< 100$\cmc, respectively, using measurements of    
[S II]$\lambda\lambda 6716/6731$ doublet line ratios and an assumed 
temperature of $10^4$\,K for the cool gas.
Using these values and uniform filling factors, the thermal pressures 
of the $4.6 \times 10^6$\,K X-ray gas and the $\sim 10^4$\,K 
optically emitting gas  appear to differ. Equality of the thermal
pressures given these assumptions would 
imply an electron density $\sim 20$\cmc for the cool component. 
However, the uncertainties are large. In addition to 
uncertainties in the emission volume and filling factors for both 
components as well as the temperature of the cool gas, the 
dependence of the [S II] 
doublet line ratios on the electron density is weak in this low
density regime such that it is not a sensitive discriminator  
between electron densities of a few $10$'s to a few $100$\cmc 
(Osterbrock 1989). Thus it is plausible that local thermal pressure
equilbrium is satisfied. Better optical as well as X-ray measurements 
would be needed to determine whether the thermal pressures are in fact 
discrepant, signaling either expansion of the optical filaments or 
the presence of non thermal pressure components. 

Three striking results emerge from our analysis. First, we see the same
extreme asymmetry in the X-ray gas as seen in other ISM tracers. X-ray
emitting gas outside the bulge is concentrated to the 
south and west of the galaxy, while little emission is seen
to the north. Second, the X-ray features 
extending out of the plane of the disk are in close 
correspondence to the H$\alpha+$[N II] features observed by Keel \&
Wehrle (1993) and Kenney \etal (1995), sharing the same angular
positions and filamentary structure. This suggests that the
filamentary features consist of a heterogeneous mixture of gas in two
or more phases, i.e. 
H$\alpha$ emitting gas at temperatures $\sim 10^4$\,K 
and X-ray gas at temperatures 
$\sim 4.6 \times 10^6$\,K, 
as well as molecular gas in filament 1 (Combes \etal 1988). 
Although arising from a different physical origin, starbursts in other
X-ray luminous galaxies, e.g. NGC 253 (Strickland \etal 2000, 2002) and M82 
(Strickland \& Stevens 2000), also show X-ray and H$\alpha$
correlations. Finally, the properties of X-ray emitting gas in the filaments,
i.e. temperature, metallicity, and density, are remarkably similar 
to those of X-ray gas in the bulge. 

Two competing scenarios, both involving a combination of tidal and 
gas interaction processes, have been proposed to explain the complex
morphology seen in NGC 4438.  One scenario uses tidal interactions
due to a close flyby of the galaxy with the Virgo cluster center 
to produce the observed stellar tidal tails (Miller 1986); while 
ram pressure and turbulent stripping of the ISM due to the motion of 
the galaxy through the intracluster medium are invoked to explain  
the trailing distribution of hot gas (Kotanyi \etal 1983). Simulations
by Miller (1986) showed that tidal forces from the close approach of
a disk galaxy to the cluster core would produce two long, oppositely 
directed stellar tails of various shapes that could include dense
folds or hooks such as seen in Figure \ref{fig:bigmap} in the 
northern stellar tidal feature in NGC 4438. Recent hydrodynamical 
simulations of ram pressure stripping in galaxies (Balsara \etal 1994;
Schulz \& Struck 2001, Acreman \etal 2003) have shown that the  
gasdynamics of ICM-ISM stripping is complex. In particular, Schulz \& 
Struck (2001) showed in 3-D simulations of ICM-ISM interactions in disk 
galaxies that the stripped gas, while displaced from the disk, may be 
``hung up'' downstream of the disk due to the increased gravitational 
influence of the dark matter halo on the displaced gas. They found 
that this might account for filamentary hydrogen gas structures 
with scales up to several galactic diameters. 
This is  consistent with the 
fact that the longest filaments in NGC 4438 are roughly twice the 
diameter of the central gas bulge. Schulz \& Struck also found that 
the temperature of the gas in the wake depended on the initial density
of the cloud.  For moderate to high densities, cooling dominated 
heating by the ICM such that much of the  gas cooled to temperatures 
$\sim 10^4$\,K. This could account for strong H$\alpha$ emitting
gas in the downstream features. Less dense clouds would be
subject to heating by the ICM to temperatures of order that of the 
cluster gas and thus would be X-ray emitting. Using the criteria from
Schulz \& Struck (2001) for NGC 4438 in Virgo, the
critical electron density between these two regimes is $\sim
0.04$\cmc, consistent with the densities observed in the X-ray
filaments. Tangled magnetic fields, either dragged out of the disk or 
produced by the dynamo effect due to turbulent motions induced by the 
stripping process, might inhibit thermal conduction between these clouds
 and delay their thermal mixing.

While qualitatively appealing, this scenario encounters several
difficulties. First, as noted by previous authors 
(e.g Combes \etal 1988; Kenney \etal 1995), ram pressure stripping by 
the Virgo ICM is roughly two orders of magnitude too weak to displace 
the molecular gas observed in filament $1$ $68''$ from the inner disk.
Second, recent simulations of ram pressure stripping 
(Abadi \etal 1999; Quilis \etal 2000; Schulz \& Struck 2001) all
agree that the process is very prompt. Even in the filamentary 
structures ``hung up'' by the halo potential, Schulz \& Struck (2001)
found that most of the gas would be stripped from the galaxy on a timescale
of a few hundred million years. Thus stripping is expected to occur
during the initial infall of the galaxy toward the cluster center 
and would most likely be complete before a close flyby of the cluster 
center could produce the observed tidal stellar tails. Furthermore, 
in the simulations, the temperature of gas shock heated by the ICM, 
with a relative velocity $\sim 1200$\kms, tends to be much higher than the 
$0.4$\,keV observed for diffuse gas in the extended filaments and the 
bulge of NGC 4438. Third, these same simulations showed 
that ram pressure also affects the structure of retained gas in the 
inner disk. In particular, this gas would suffer a bulk displacement 
from the dynamical center of the galaxy. Subsequent radial compression of
the inner disk gas would result in the onset of gravitational
instabilities and a ring of enhanced density and star formation on 
timescales of a few disk dynamical times, 
i.e. $\sim 1 - 3 \times 10^8$\,yrs. These effects are not 
seen in the central region of NGC 4438. This suggests that the
duration of gas-gas interactions in NGC 4438 was short, much less 
than the $\sim 10^8$\,yr rotation period for gas in the bulge, so
that gravitational instabilities would not have had time to 
develop and the effects of any disturbance of the retained gas in the 
central regions could have been erased. Finally while ram pressure
caused by the motion of the galaxy through the ICM would produce 
characteristic asymmetries between the leading bow-shaped edge of 
the galaxy and the trailing gas wake, most simulations show 
symmetry of the wake about the central axis of the galaxy along its 
direction of motion. Thus it is 
more difficult to explain the paucity of stripped gas to the 
northwest, relative to that in the south. In addition, as noted by Kenney
\etal (1995), if cluster tidal forces and ICM stripping are
responsible for the observed morphology, it seems unlikely, given 
the number of other galaxies near the Virgo cluster core, 
 that NGC 4438 alone should show such severe damage.

A more natural explanation for the observed features in NGC 4438 is
that we are seeing the aftermath of a relatively high velocity,
off-axis collision with nearby NGC 4435. 
Combes \etal (1988) were able to reproduce the observed stellar tidal 
tails and the northwest stellar fold in simulations of a retrograde 
collision $10^8$ years ago between NGC 4435 and NGC 4438 with relative
velocity $\sim 900$\kms and point of closest approach $\sim 5$\,kpc 
(about twice the radius of the X-ray 
emission region in the bulge) to the north of NGC 4438's optical
center. Such a collision velocity is reasonable, given the observed 
$730$\kms line-of-sight relative velocity between the two galaxies 
and their close proximity in projection. The observed asymmetries 
in the tidal distortions are also
easier to explain in the context of galaxy-galaxy collisions.
The gravitational potential from the less massive NGC 4435 changes 
significantly over galaxy-sized scales, such that the tidal forces due to
the collision would be significantly different on the near and far side of 
NGC 4438 from the collision point. This would produce strong 
distortions in the stellar distribution to the northeast and northwest
closest to the collision path and a small angular displacement from
the disk direction for the southwest stellar ridge (see 
Figure \ref{fig:bigmap}). In contrast, cluster induced tidal forces do
not change significantly over galaxy scales and thus tend to produce
more symmetrical tidal features (Byrd \& Valtonen 1990). 

Similarly the asymmetric distribution of the extended X-ray filaments
and their close correlation with the H$\alpha+$[N II] filaments argue
for ISM-ISM interactions in an off-axis galaxy-galaxy collision to the
north as their common origin (Kenney \etal 1995). In hydrodynamical 
simulations of ISM-ISM interactions in a single phase medium in
off-axis galaxy collisions, 
M{\"u}ller \etal (1989) found that the resulting properties 
of the displaced ISM strongly depended on the relative velocity
between the two galaxies. They found that due to 
the lower kinetic energy in galaxy-galaxy collisions with relative
velocities $v \lesssim 1000$\kms, such as proposed here, 
cooling within the shock fronts could not be neglected. 
The maximum temperatures found for the disturbed gas in these
simulations were only $\sim$ a few times $10^6$\,K, consistent with 
the low ($4.6 \times 10^6$\,keV) X-ray temperatures
 we observe in the bulge and extended filaments 
in NGC 4438.  They also found, depending on the masses of the 
colliding partners and the impact parameter of the collision, that 
the displaced gas could remain bound to the galaxy. The low 
velocities observed for gas in the extended optical 
filaments in NGC 4438 (Kenney \etal 1995; Chemin \etal 2003) suggests 
that this may be the case, with gas in the 
filaments still bound to NGC 4438 and in the process of falling back 
onto the disk. The timescale over which ISM-ISM interactions are 
effective in this collision ($\lesssim 10^7$\,yr depending on the
scale heights of the colliding gas; Kenney \etal 1995) is small 
compared to the elapsed time since the collision. This explains the 
relatively undisturbed appearance of gas within the central bulge
since, in contrast to the case of ICM-ISM interactions, it will have 
had a number of dynamical time periods since the gas interactions 
have ceased to recover. An off-axis collision of two galaxies
naturally explains the observed north-south asymmetry seen in the
gas. Without ISM-ISM interactions, gas should be tidally disrupted
in roughly the same tidal streams as the stars. However, due to the
stronger ISM-ISM interactions to the north near the point of closest
approach between the galaxies' centers ( as well as the stronger 
tidal interactions as the centers pass each other), gas outside 
the bulge is efficiently swept  from the region, explaining the 
lack of both H$\alpha$ and X-ray emission there. 
ISM-ISM interactions are expected to be weaker to the
south of the bulge, perhaps explaining the small ($\lesssim 8''.7$) 
displacement of filament 3 from the southwestern tidal stellar ridge.

While suggestive, we do not expect the symmetric, single phase 
hydrodynamic simulations of M{\"u}ller \etal (1989) to reproduce in 
detail the properties of a multi-phase ISM-ISM collision. 
Instead, we expect that (as in the ICM-ISM 
simulations of Schulz \& Stuck, 2001) the balance between heating and
cooling in clouds of different densities would profoundly affect the 
properties of the gas we now observe. Clouds above the now higher
($\gtrsim 0.1$\cmc) critical density cool efficiently to $10^4$\,K, 
accounting for the strong H$\alpha$ emission in the filaments. 
This is consistent with the higher densities 
now
observed in the H$\alpha$ gas (Kenney \etal 1995).
X-ray emissions come from gas streams of initially lower density
that are  heated by the ISM-ISM interaction, also consistent with the 
lower densities 
we measure for this component.
Due to the low metallicity as well as low temperatures found for this
gas in our spectral fit, the cooling time for this gas
($\sim 4 - 7 \times 10^8$\,yrs) is much longer than the 
$10^8$\,yrs since the collision. Thus we do not expect that much of 
this gas would have cooled. 

ISM-ISM ram pressures in a  collision between
NGC 4438 and NGC 4435 are expected to be $1$ to $3$ orders of
magnitude larger than the corresponding ISM-ICM ram pressures from
 the motion of NGC 4438 through the Virgo cluster gas making it 
easier to explain how  ISM-ISM interactions in concert with tidal
forces could displace dense molecular
gas $68''$ out from the center of the disk in  filament $1$ 
(Combes \etal 1988).
The close correspondence of the X-ray features seen in our high 
angular resolution Chandra images of NGC 4438 to those
seen in H$\alpha$ emission by Kenney \etal (1995)
suggests that these two heterogeneous gas
phases coexist in the same filamentary structures. Although not
required, magnetic fields may also have been dragged out of the disk
and tangled due to turbulence produced in the collsion. Such a network
of magnetic fields would both link the different gas phases and delay 
their thermal mixing. In such a case 
the forces binding multiphase gas to the disk would 
depend upon an average surface gas density over some larger region, 
rather than the local surface gas density within a particular cloud.
 Since these average densities typically tend to
be an order of magnitude lower than the local density found within 
dense molecular clouds, this would further enhance the ability of 
ISM-ISM interactions to drag dense, molecular gas out of the disk 
along with the less dense ISM material during the collision (Kenney
\etal 1995).

\subsection{NGC 4438: The Nuclear Emission Region}
\label{sec:core}

In Figure \ref{fig:core} we focus on the structure of the inner nuclear 
region of NGC 4438. 
To obtain a sharper Chandra image of the innermost nuclear region, we 
reprocessed our data to remove the standard pixel randomization. 
In the left panel of Figure \ref{fig:core}, we show the resulting
soft band ($0.3-2$~keV) image of the nuclear region smoothed with 
a $0''.5$ Gaussian. 
The X-ray emission can be separated into a
dominant elongated northwest feature,  
containing the peak X-ray emission and
the nucleus, and a weaker feature 
located $\sim 9''$ to the southeast.
In the middle panel, we show the $0''.1$ resolution
HST H$\alpha$ ($\lambda 6563$) image taken by Kenney \& Yale (2002)
matched to the X-ray image using WCS coordinates.
Strong correlations exist between these nuclear X-ray features and the 
outflow bubbles observed in H$\alpha+$[N II] with HST by 
Kenney \& Yale (2002).
The northwestern X-ray feature corresponds to the northwest 
outflow bubble seen in the middle panel of Figure \ref{fig:core}  
with the peak X-ray emission near the southeastern tip, 
$\lesssim 0''.7$ from 
the location of the compact optical nucleus. 
While not readily visible in the H$\alpha$
image alone, the X-ray feature  $\sim 9''$ to the southeast of the 
nucleus corresponds to 
the southeastern outflow shell identified by Kenney \& Yale (2002) in
the continuum subtracted sum of H$\alpha +$[N II] HST images. This
X-ray emission resolves into a ring-like pattern similar to
the less organized, patchy emission observed in 
H$\alpha+$[N II]. We, however, observe no X-ray emission 
(L$_{\rm X} \lesssim 7.4 \times 10^{36}$\ergs) 
from the location of the strong HII regions seen $8''$ south of the
nucleus in the middle panel
of Figure \ref{fig:core}. 
 
In the right  panel of Figure \ref{fig:core} we superpose Chandra
X-ray contours from a lower resolution $1''$ Gaussian smoothed, 
background subtracted, exposure corrected Chandra image of the nuclear
emission region in the $0.3-2$~keV energy band on a $\sim 5''$ resolution 
$20$\,cm FIRST Survey map of the core region 
(see http://sundog.stsci.edu/). 
The X-ray emission is also strongly 
correlated to the radio continuum 
emission with the northwestern  and southeastern nuclear X-ray peaks 
coinciding with radio continuum peaks A and B reported by 
Hummel \& Saikia (1991). These correlations between X-ray, H$\alpha$
and radio features suggest a common origin for the emissions in shocks
and postshock outflow gas, after the gas impacts the surrounding ISM.

\begin{figure}[t]
\begin{center}
\resizebox{10.0cm}{!}{\includegraphics{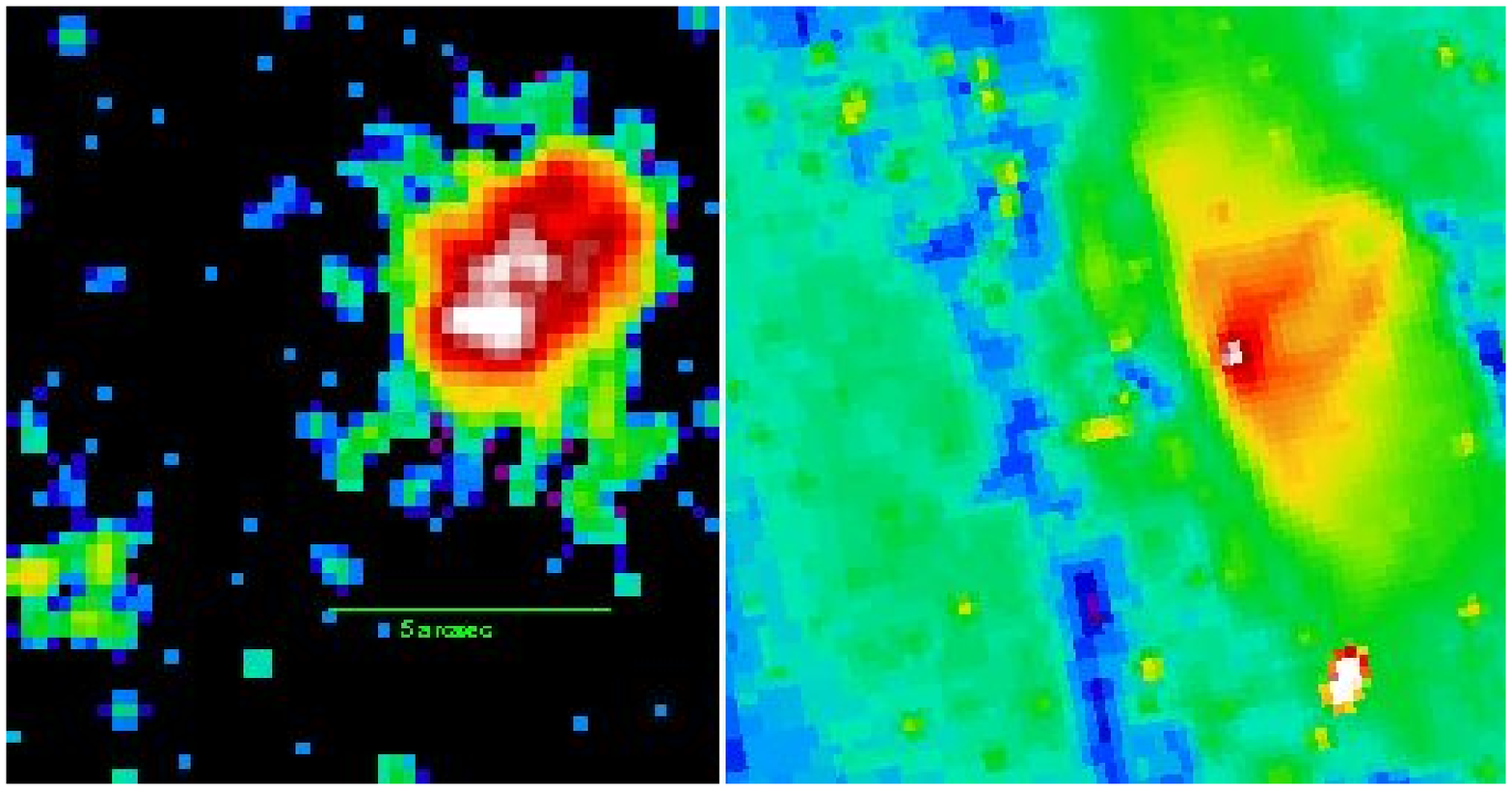}}
\resizebox{5.0cm}{!}{\includegraphics{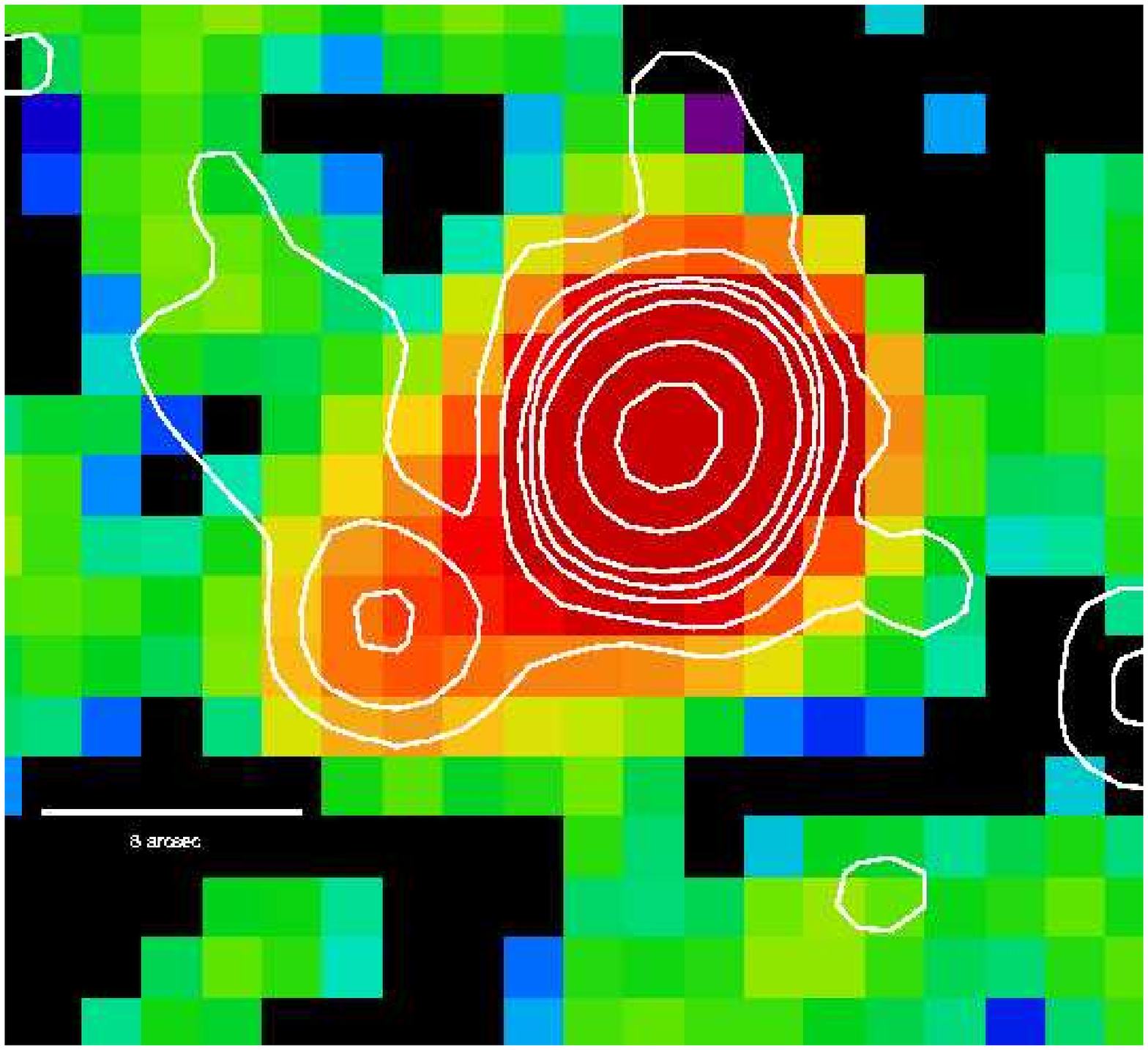}}\\%
\caption{(left) $14'' \times 14''$ nuclear region in the 
$0.3 - 2$~keV energy band with
pixel randomization removed and smoothed with a $\sigma=0''.5$
Gaussian. The scale bar is $5''$. (middle) HST image ($\lambda6563$,
Kenney \& Yale,2002) of the same region matched in WCS coordinates to
the X-ray image on the left. 
(right) VLA First image of the $32'' \times 32''$ field surrounding the 
nuclear region of NGC 4438, superposed with contours from a $\sigma=1''$
Gaussian smoothed, background subtracted, and exposure corrected
Chandra X-ray image of the nuclear region. Contour levels are 
$0.4$, $0.8$, $1.6$, $2.5$, $4$, $12$, $25 \times 10^{-7}$\expunit. 
The scale bar in the right panel is $8''$. In all three images North
is up and East is to the left.
\label{fig:core}}
\end{center}
\end{figure}

We used a $5''$ circular 
region centered at 
($12^h 27^m 45.6^s$, $+13^{\circ} 00' 32.8''$) to study the 
spectral properties of the northwest
outflow region, including the nuclear source.
We chose a local background
annulus with inner radius $7''.4$ and 
outer radius $20''.7$ centered on the source region 
to subtract the backgrounds due to the outer galactic diffuse emission
and Virgo cluster gas. After background subtraction,
the nuclear (northwest outflow) region had $1120 \pm 34$ net source counts
in the $0.3-5$~keV energy band used in the spectral fits.
We again fit spectra using XSPEC with a minimum of $20$
counts per spectral bin.  
We used four different models each with a variable absorbing column, i.e.
a single component power law, single
component MEKAL, variable MEKAL, and power law plus MEKAL model,
to probe the properties of the nuclear outflow gas and 
a possible nuclear source. 
These fits are summarized in Table \ref{tab:nucspec}.
\begin{deluxetable}{cccccc}
\tablewidth{0pc}
\tablecaption{Northwest Nuclear Outflow Region Spectra\label{tab:nucspec}}
\tablehead{\colhead{Model}  & \colhead{PL} &\colhead{MEKAL} &\colhead{VMEKAL}
&\colhead{ PL + MEKAL}} 
\startdata
N$_{\rm H}^a$($10^{22}$\cms) &-- &$0.20^{+0.04}_{-0.05}$ 
&$0.19^{+0.05}_{-0.04}$ &$0.19^{+0.10}_{-0.04}$ \\
$kT$ (kev)   &--  &$0.60^{+0.04}_{-0.04}$  &$0.61^{+0.04}_{-0.05}$ 
& $0.58^{+0.04}_{-0.10}$ \\
$A$ ($\Zs$)  &--  &$0.14^{+0.06}_{-0.03}$ &$0.11^{+0.10}_{-0.04}$ & $0.14^f$\\
$A\alpha$ ($\Zs$)  &-- &-- &$0.15^{+0.21}_{-0.10}$ & --  \\
N$_{\rm H}^b$($10^{22}$\cms)&$1.13$&--  &-- & $2.9^{+3.1}_{-2.0}$\\
$\Gamma$ & $9.0$  &-- &-- &$2.0^f$ \\
$\chi^2/{\rm dof}$   &$182/40$ &$62/39$ & $61.5/38$ & $39.6/38$ \\
\enddata
\tablecomments{ Model components are power law (PL), MEKAL, and
variable MEKAL (VMEKAL). Fits include the nuclear source. N$_H^a$  
(N$_H^b)$ are hydrogen column densities for the thermal (power law)
components. Superscript $f$ specifies that the parameter is held fixed.
}
\end{deluxetable}
A single, absorbed power law is unable to describe the
spectrum, yielding an unphysical photon index $\Gamma=9.0$ and
unacceptable $\chi^2$.
An improved, but still unacceptable, fit (see Table \ref{tab:nucspec}) 
is obtained for a highly
absorbed single temperature MEKAL model. 
The fit is not significantly improved 
by using a variable MEKAL model and allowing the alpha and
non-alpha element abundances to vary independently. 
 A good fit is obtained using two components, a 
highly absorbed power law to model the nucleus and an 
absorbed, single temperature MEKAL component to model emission from
the outflow bubble. Due to our limited statistics, we select the 
abundance as that from the single thermal fit and consistent with
nuclear outflow gas in M51 (Terashima \& Wilson 2001). 
We fix the photon index at $2.0$, consistent with an absorbed power
law fit to the data above $2$\,keV and with previous ASCA measurements
of this system (Terashima \etal 2002).
We then allow the gas temperature and both absorbing columns to
vary. We find the data are well fit with a gas temperature and absorbing 
column for the outflow component of $kT = 0.58^{+0.04}_{-0.10}$\,keV and
$n_H = 1.9^{+1.0}_{-0.4} \times 10^{21}$\cms,
consistent at the $90 \%$ confidence level with the MEKAL and VMEKAL
fits listed in Table \ref{tab:nucspec}, 
and a moderately 
obscured ($n_H \sim 2.9^{+3.1}_{-2.0} \times 10^{22}$\cms)
power law component for the nuclear source.
We show the spectrum and this best fit absorbed power law plus 
absorbed MEKAL model in Figure \ref{fig:n4438nucspec}.


\begin{figure}[t]
\epsscale{0.5}
\begin{center}
\epsfig{file=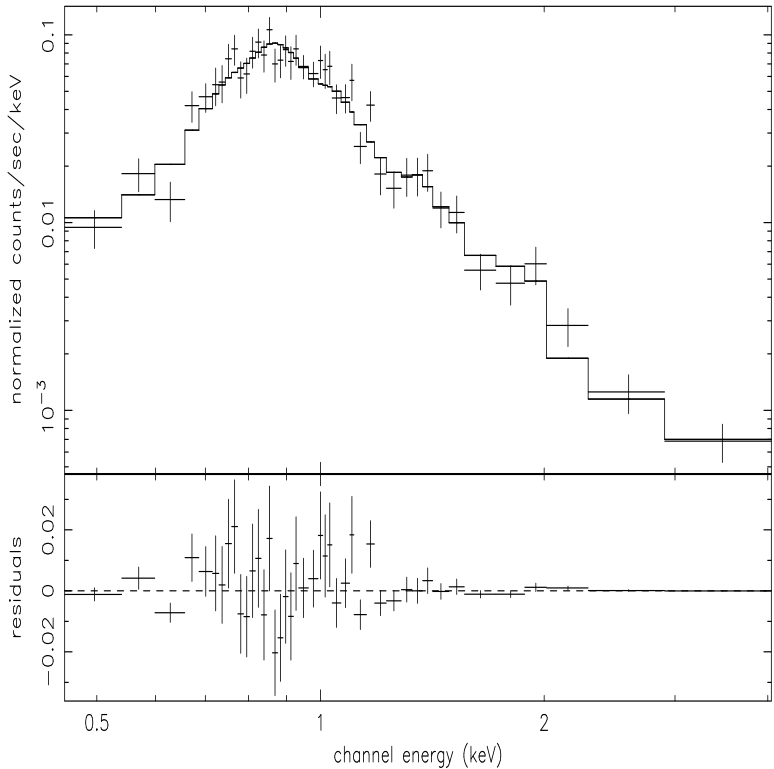,height=4in,width=4in,angle=270}
\caption{Spectrum and best fit absorbed power law plus absorbed
MEKAL model for the nucleus and northwest nuclear outflow region
of NGC 4438.
\label{fig:n4438nucspec}}
\end{center}
\end{figure}

\begin{deluxetable}{ccccc}
\tabletypesize{\footnotesize}
\tablewidth{0pc}
\tablecaption{Nuclear Region Observed Fluxes and Intrinsic Luminosities\label{tab:nuclum}} 
\tablehead{
\colhead{Component}  &\colhead{ flux ($0.3-2.0$\,keV) } 
&\colhead{L$_{\rm X}$ ($0.3-2.0$\,keV)} 
& \colhead{flux ($2.0-10.0$\,keV)} &\colhead{L$_{\rm X}$ ($2.0-10.0$\,keV)}\\
 &($10^{-13}$\ergscm)& ($10^{39}$\ergscm) &($10^{-13}$\ergscm)
 & ($10^{39}$\ergscm) } 
\startdata
 thermal & $1.41$ & $10.2$ & $0.08$ &$0.26$ \\
 power law & $0.03$ & $2.4$ & $0.51$ & $2.04$ \\
 total & $1.43$ & $12.6$ & $0.59$ & $2.3$ \\
\enddata
\tablecomments{Values above assume the 
absorbed power law plus MEKAL model given in 
Table \protect\ref{tab:nucspec}. Fluxes are observed. Luminosities are
corrected for absorption. 
}
\end{deluxetable}

Fluxes and luminosities for the northwest outflow region as a whole
and for the thermal (outflow gas) and power law (nuclear source)
components separately are listed in Table \ref{tab:nuclum} for the 
soft ($0.3-2$\,keV) and hard ($2-10$\,keV) energy bands. 
The soft band, where we have
the most sensitivity, is dominated by thermal emission from the
outflow gas, whose soft band observed flux and absorption-corrected 
luminosity are 
$1.4 \times 10^{-13}$\ergscm and $1.0 \times 10^{40}$\ergs. If we
assume that this gas uniformly fills a sphere of radius $5''$ 
(the size of the spectral extraction region), we infer a mean electron
number density and mass of hot gas in the northwest outflow region of
$<n_e^2>^{1/2} \sim 0.5$\cmc and $3.5 \times 10^6\Ms$. The thermal
pressure of the $0.6$\,keV X-ray gas at this density is comparable to 
that in the $10^4$\,K gas ($n_e \sim 300 - 400$\cmc, Kenney \& Yale
2002), consistent within the large uncertainties with local pressure 
equilibrium between these two components.
 Using equation \ref{eq:cool} and a cooling function 
$\Lambda \sim 6.6 \times 10^{-24}$\,erg\,cm$^3$\,s$^{-1}$, the
cooling time for the X-ray outflow gas is only
$\sim 3 \times 10^7$\,yr. 
Thus heating of the outflow gas must have occurred quite
recently. 

The steep spectral index and location of the hard X-ray emission 
at the center of the galaxy are suggestive of an AGN. 
While the detection of 
the Fe K line would clearly point to an AGN as the nuclear source, in 
the absence of any detectable emission above $3$\,keV, we are unable
to measure, or meaningfully bound, parameters for the Fe K line. 
However, producing the steep, hard power law
component from X-ray binaries is unlikely because the $2 - 10$\,keV 
luminosity for the power law component, 
$L_{\rm X} \sim 2.0 \times 10^{39}$\ergs (see Table \ref{tab:nuclum}),
 is a factor $36$ larger
than expected for HMXB's (Gilfanov \etal 2003) 
given the upper limit on the nuclear star 
formation rate of $0.1\,\Ms$yr$^{-1}$  
determined from radio continuum data (Hummel \& Saikia 1991; 
Kenney \& Yale 2002). We also note that the low AGN luminosity and 
the steep power law slope are consistent with an ADAF model (Narayan
\& Yi 1995) where X-rays are produced by inverse Compton scattering of
synchrotron photons off flow electrons (See, for example, the analysis of 
the nuclear regions in M87, Di Matteo \etal 2003, and in NGC 4594, 
Pelligrini \etal 2003).  

The fitted  column densities for the nucleus and outflow
gas are consistent with optical extinction measurements of the
region. Ho \etal (1997b) used H$\alpha$/H$\beta$ line ratios  in a
$2'' \times 4''$ aperture about the nucleus 
(thus including contributions from the northwest outflow shell) to 
infer a weighted, mean intrinsic color excess $E(B-V)$ for this 
region of $0.61$. Assuming a normal dust to gas ratio, such that
\begin{equation} 
\frac{n_{\rm H}}{E(B-V)} = 5.28 \times 10^{21}\,{\rm cm}^{-2}
\label{eq:dust}
\end{equation}
 (Binney \& Merrifield 1998, eq. 3.65), this color excess implies 
a mean intrinsic hydrogen column 
$n_{\rm H} \sim 3.2  \times 10^{21}$\cms (total hydrogen
column $n_{\rm H} \sim 3.5  \times 10^{21}$\cms). This column is 
consistent with, although marginally higher than, our fitted value for
absorption of the outflow (thermal) component and a factor $3 - 17$ 
(given our large uncertainties) less than our fit to the absorption
of the nuclear (power law) component. This is to be expected 
since, as pointed out by Kenney \& Yale (2002), 
the observed H$\alpha+$[N II] flux from the
northwest outflow shell is $7$ times that of the nucleus   
and the determination of the extinction corrections by Ho \etal (1997b) 
included a significant contribution from outflow gas as well as the
nucleus. Thus our measured values are reasonable. Similarly
the temperature we find for the outflow gas 
($0.58^{+0.04}_{-0.10}$\,keV) is similar to temperatures measured for 
gas surrounding nuclei in other active galaxies such as M51 
($0.5$\,keV,Terashima \& Wilson 2001)
and NGC 4594 ($0.6$\,keV, Pellegrini \etal 2003). 

There are too few counts in the southeast outflow shell 
($35 \pm 6$ in a circle of radius $1''.7$) 
to obtain a separate spectrum for that region. If we assume that the 
southeast outflow shell has the same temperature and absorbing column
as the northwest outflow gas, we estimate its observed $0.3-2$\, keV flux and
intrinsic luminosity to be $\sim 4.8 \times 10^{-15}$\ergscm 
and $\sim 3.4 \times 10^{38}$\ergs. This also implies an average electron 
density and cooling time of $<n_e^2>^{1/2} \sim 0.5$\cmc and
 $t_{cool} \sim 3 \times 10^7$\,yr.
Because of the low luminosity of the southeast outflow region, it is more
difficult to define the bubble geometry and infer the X-ray gas mass. 
We estimate the mass of hot gas in this structure to be at least 
$1.3 \times 10^5\Ms$. However, it could possibly be as high as that found
in the northwest outflow structure.

Kenney \& Yale (2002) found that the optical line emission responsible
for the LINER classification of NGC 4438 originated from shocked gas 
at working surfaces, where outflow gas interacts with the
surrounding interstellar medium. If shocks also are responsible for 
heating the X-ray gas, we can use the outflow gas temperature to determine
the bulk outflow velocity. The expected post-shock temperature 
$T_s$ of a fully ionized gas
with $10\%$ helium to hydrogen in number density is (Hollenbach \& 
McKee 1979) 
\begin{equation}
T_s = 1.4 \times 10^{5} V^2_{7}
\label{eq:shocks}
\end{equation}
where $V_{7}$ is the shock velocity in units of $10^7$\,cm\,s$^{-1}$. 
Therefore the 
shock velocity needed to heat gas to $0.5 - 0.6$\,keV in the nuclear
region is $600 - 700$\kms. Kenney \& Yale (2002) find evidence from 
[N II] and [S II] line widths for line of sight gas velocities 
$\sim 300$\kms. The higher shock velocity in the X-ray emitting gas 
compared to that calculated for the optical is expected since the 
two gas phases have different densities. 

The sizes of the outflow regions and their cooling times are
consistent with the outflow having traveled from the nucleus in the 
recent past. The northwest outflow bubble has a maximum extent of 
$\lesssim 4''.7$ ($360$\,pc). Even assuming the conservative lower
limit to the outflow velocity ($300$\kms) measured by Kenney \& Yale (2002), 
the outflow could have originated in the
nucleus $1.2 \times 10^6$ years ago, much less than the calculated cooling
time of
$\sim 3 \times 10^7$\,yr 
for gas in this bubble. The southeastern outflow bubble is twice as far
from the nucleus ($9''.5$ or $\sim 730$\,pc) and much fainter, consistent
with it being an older structure with the nuclear outburst occurring 
twice as long ago 
as that which formed the northwestern outflow. 
This is still much less than the cooling time 
$\sim 3 \times 10^7$\,yr 
for the gas. 
It is also of order the synchrotron lifetime for 
electrons in the corresponding radio feature found by 
Hummel \& Saikia (1991), 
thus confirming the need for high outflow velocities $\sim 700$\kms and 
explaining the weakness of the radio source. 

An alternative explanation for the different extent and brightness of
the two outflow bubbles could be that the outflows represent bubbles of the 
same age but in different stages of evolution. This might be the case
for outflows propagating through ISM of different densities.
The SE outflow structure propagating into less 
dense material would then be assumed to 
be in the adiabatic or free-expanding phase such that it would radiate
only weakly; while the NW outflow
shell would be in the superbubble stage of evolution and would 
radiate more efficiently due to its rapid deceleration by 
denser ISM material (Kenney \& Yale 2002).  

In either case, the energetics of the outflow gas provides insight
into the nature of the nuclear power source. The amount of hot, ionized gas
in the northwest outflow region 
($3.5 \times 10^6\,\Ms$) is more than a factor of $30$ greater than
the lower limit placed on the gas by the H$\alpha$ measurements alone
(Kenney \& Yale 2002). Assuming the measured lower limit on the
outflow velocity of $300$\kms, the kinetic energy contained in this
gas is $\sim 3 \times 10^{54}$\,ergs, close to the kinetic energy
found in the ionized gas outflow of NGC 3079 ($2 \times
10^{54}$\,ergs, Veilleux \etal 1994; Cecil \etal 2002) and greater than the 
$\sim 10^{53}$\,ergs 
lower limit estimated by Kenney \& Yale (2002) for NGC 4438. 
If we assume that the 
nuclear source is a compact starburst, this amount of kinetic
energy would require the energy output of more than $3000$ supernova
from a region $\lesssim 77$\,pc in radius and, following the arguments
of Kenney \& Yale (2002), a star formation rate 
$\gtrsim 0.3\,\Ms$yr$^{-1}$ within the inner $77$\,pc of NGC 4438. 
This star formation rate 
is  above the upper limits on star formation for a nuclear starburst in this
galaxy, derived from measurements of the H$\alpha$ emission
($0.05\,\Ms$yr$^{-1}$) and 
the radio continuum power ($0.1\,\Ms$yr$^{-1}$)
(Kenney \& Yale 2002). If we use the more realistic outflow velocity
$\gtrsim 600$\kms from the shock heating analysis, the kinetic energy 
in the outflow gas could be a factor of $4$ higher requiring a star 
formation rate in excess of $1\Ms$yr$^{-1}$. Such a large kinetic 
energy in the outflow gas coupled with the measured low values for the
nuclear star formation rate favors a central AGN, either alone or 
in addition to nuclear star forming activity,
as the driver of the outflows. It is perhaps interesting to note 
that the kinetic pressure from this outflow 
($1-4 \times 10^{-9}$\,dyne\,cm$^{-2}$) is  comparable to the 
thermal pressure ($\sim 10^{-9}$\,dyne\,cm$^{-2}$) of the $0.6$\,keV
gas. An AGN interpretation for the nuclear 
source also could explain the high degree 
of collimation in the southeastern bubble, the enhanced radio
brightness observed at the ends of the outflow shells, and the
weak $2050$\kms broad-line component observed in the nuclear spectrum by 
Ho \etal (1997c), but may require the AGN jets to be subrelativistic 
and thus not easily observed (Kenney \& Yale 2002; Bicknell \etal
1998). 

Our best fit spectral model for the 
northwest outflow region does include a weak, power law component that could
be attributed to an AGN nuclear source.
However, this power law component is heavily absorbed, contributing 
predominately to the high energy tail of the observed spectrum. The 
power law plus MEKAL spectral model shown in Table \ref{tab:nucspec} 
predicts a $2.0 - 10$\,keV intrinsic luminosity for the nuclear source 
of $2 \times 10^{39}$\ergs. For this spectral index, the bolometric 
luminosity is about a factor of $10$ larger 
(see, e.g. Pellegrini \etal 2003), 
$L_{\rm bol} \sim 2 \times 10^{40}$\ergs. However, due to our limited 
statistics at high energies, the errors, particularly on the 
absorbing column, are quite large. This
translates into $\sim 60\%$ uncertainty in the intrinsic 
luminosity of the nuclear source.
Using the bulge luminosity (Kenney \etal 1995) and the 
correlation between bulge luminosity and black hole mass 
(Ferrarese \& Merrit 2000; Gebhardt \etal 2000), Kenney \etal (1995) 
inferred a black hole mass of $5 \times 10^7\Ms$ for NGC
4438. However, the ratio of the bolometric luminosity of the 
nuclear source to the Eddington luminosity is extremely small,
$L_{\rm bol}/L_{\rm Edd} \sim 3 \times 10^{-6}$. Such  
sub-Eddington accretion appears to be common in most
nearby galaxies (Roberts \& Warwick 2000). This ratio is similar to that
found in NGC 4594 where the weakness of the source is attributed to
radiatively inefficient or unsteady accretion (Pellegrini \etal
2003). Unsteady accretion onto a
black hole in NGC 4438 might both account for the nuclear outflows and
reconcile the energy budget for the flows, since the nuclear
AGN would be episodic and could have been more energetic in the recent past. 

In the likelihood that  the nuclear source is an AGN, then 
the $2-10$\,keV X-ray and 
H$\alpha$ luminosities should obey the $L_{\rm X} - L_{H\alpha}$ 
relation
\begin{equation} 
\log L_{\rm X}=(1.11 \pm 0.054)\log L_{H\alpha}-(3.50 \pm 2.27)
\label{eq:lxla}
\end{equation}
 (Ho \etal 2001; see also Terishima \etal
2000 for a slightly different fit). 
Previous measurements of the 
$L_{\rm X}/L_{H\alpha}$ ratio in NGC 4438 were anomalously low, 
in part due to the inability to separate the contribution of the nuclear source
from that of the outflow regions. Using high resolution HST
measurements to resolve the nucleus and extinction corrections derived
from measurements of H$\alpha$/H$\beta$ line ratios by Ho \etal (1997b), 
Kenney \& Yale (2002) found the unabsorbed H$\alpha$ luminosity of
the nuclear source was 
$L_{H\alpha} \sim 5.5 \times 10^{39}$\ergs. Using the $2 - 10$\,keV
unabsorbed luminosity for the power law component in our best spectral
fit for the X-ray luminosity of the nuclear source, 
$L_{\rm X} \sim 2 \times 10^{39}$\ergs, gives 
$\log (L_{\rm X}/L_{H\alpha}^{1.11}) =-4.8$, 
consistent with Equation \ref{eq:lxla}. However, given the large 
uncertainties in the X-ray luminosity for the power law component, 
due to our limited statistics, and also in the 
extinction corrections for the H$\alpha$ emission, 
the $L_{\rm X}/L_{H\alpha}$ ratio could be much lower.

\subsection{NGC 4435}
\label{sec:n4435}

\begin{figure}[t]
\epsscale{0.5}
\epsfig{file=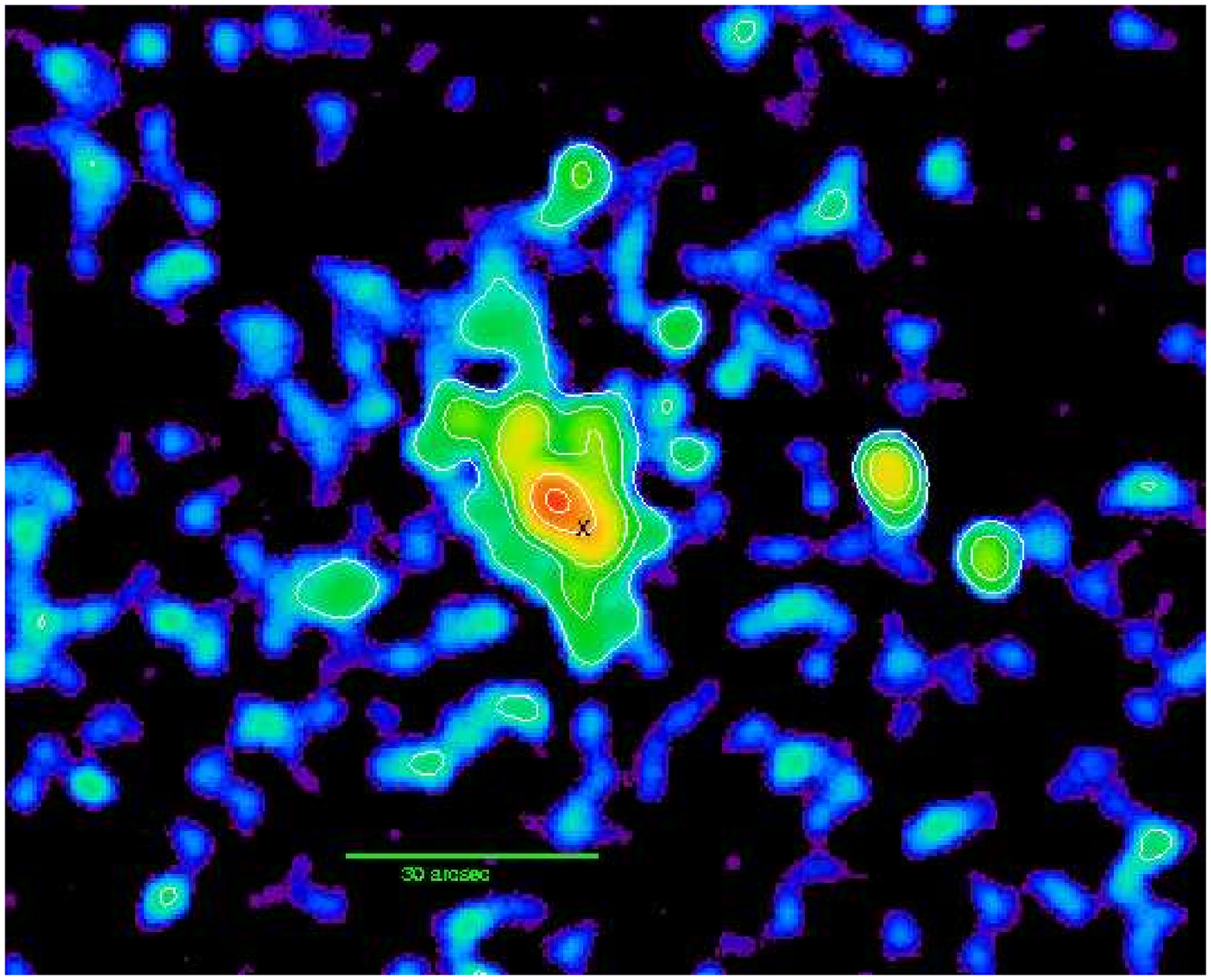,height=3in,width=3in}\hspace{0.3cm}
\epsfig{file=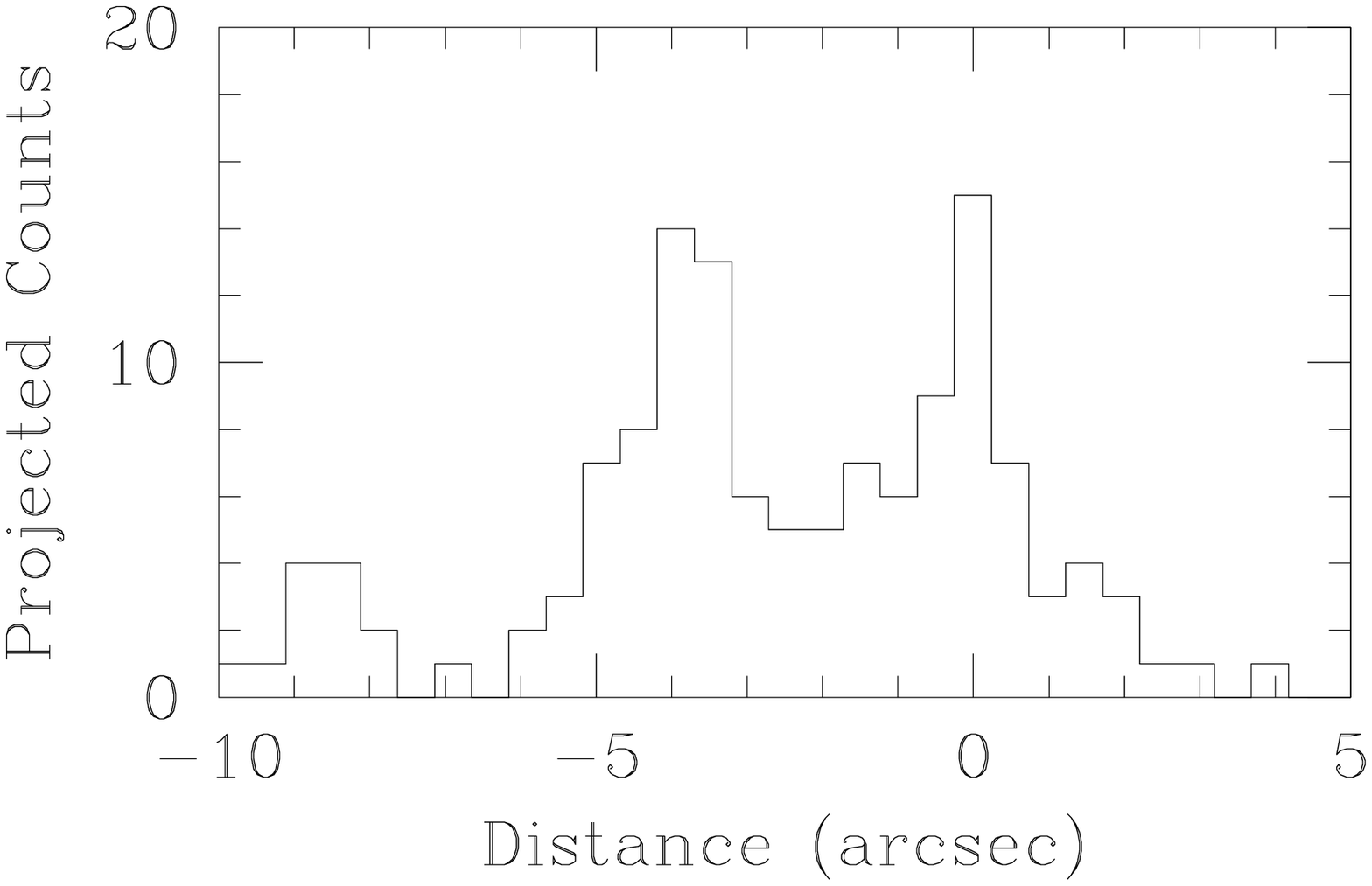,height=3in,width=4in}
\caption{(left)Chandra image of NGC 4435 in the $0.3-2$~keV energy
band. The image has been smoothed with a $\sigma=2''$ Gaussian,
background subtracted and exposure corrected. Contours are
$1.24$, $2.48$, $4.13$, $12.4$, $20.7$ in units of $10^{-8}$\expunit.
X locates the peak of the $2-6$\,keV hard band emission and the 
optical center of the galaxy. North is up and East is to the left.
The horizontal scale bar is $30''$ ($2.3$\,kpc). (right) Projection 
of the $0.3-6$\,keV image in a box of width $4''.7$ centered on the 
hard emission peak (X) onto a line oriented $45^{\circ}$ 
counterclockwise from North connecting the hard and soft emission
peaks that shows their $4''.1$ separation. For comparison
the radius for $50\%$ encircled energy of the ACIS point spread
function at this position ($3'.9$ off-axis) is $2.73 \pm 0.32$ arc seconds.
 }
\label{fig:n4435}
\end{figure}

Figure \ref{fig:n4435} shows the central X-ray emission from NGC 4435.
The X-ray morphology of this galaxy is asymmetric and knot-like. 
The dominant emission in the soft band ($0.3-2$\,keV) is centered at 
$12^h27^m40.72^s$ $13^{\circ}04'47.4''$; while the hard band emission 
($2 - 6$\,keV) is centered at $12^h27^m40.51^s$ $13^{\circ}04'44.6''$
and is coincident with a VLA FIRST radio source 
($12^h27^m40.54^s$ $13^{\circ}04'44.9''$) of integrated flux 
$2.16$\,mJy that lies at the galaxy nucleus (Argyle \& Clements 1990).  
We detect emission to a radius of $35''$ ($2.7$\,kpc). Excluding the
individual sources we find $250\pm23$ net counts (after background 
subtraction) to a radius of $35''$. 

In order to measure the gas density and mass in the core of NGC 4435, 
we need to determine the spectral parameters of the emission.
We analyzed the total emission from the galaxy (within $32''$ of the 
nucleus). Background is subtracted using
a local annular region of inner radius $36''$ and outer radius $72''$,
 also centered on the nucleus. 
None of the absorbed single component
models we considered, i.e power law, thermal bremstrahlung, or 
MEKAL plasma model, gave an acceptable fit to the data. 
The data were 
well described by an absorbed two component model, a thermal MEKAL
plasma component consistent with the presence of diffuse ISM gas plus a 
power law component to model the hard emission. Due to our limited 
statistics we were not able to allow all the parameters in the model 
to freely vary. Instead we fix the absorbing column and abundance at 
reasonable values, allowing the temperature, photon index, and both 
component normalizations to vary. We then consider the sensitivity of
our results to these choices.
We initially fix the absorbing column at the Galactic value 
($2.64 \times 10^{20}$\cms) and the MEKAL model abundance at
$0.1\,\Zs$, consistent with that found in NGC 4438.
We find
a temperature for the diffuse gas of 
 $kT=0.27^{+0.09}_{-0.07}$\,keV and photon index 
$\Gamma=1.22^{+0.26}_{-0.25}$($\chi^2/{\rm dof}=22.2/19$).
The observed fluxes for the sum of both components in the  soft
($0.3-2$\,keV) and hard ($2-10$\,keV) bands are 
$7.0 \times 10^{-14}$\ergscm and $1.25 \times 10^{-13}$\ergscm for this
model, corresponding to intrinsic soft and hard X-ray luminosities of
$2.6 \times 10^{39}$\ergs and 
$3.9 \times 10^{39}$\ergs, respectively, at a distance of $16.1$\,Mpc.
The thermal component only has a significant contribution in the soft
 band with  $0.3-2$\,keV intrinsic luminosity of 
$1.4 \times 10^{39}$\ergs; while the non-thermal soft ($0.3-2$\,keV)  and 
hard ($2-10$\,keV) band intrinsic luminosities are 
$1.2 \times 10^{39}$\ergs and $3.9 \times 10^{39}$\ergs.
Assuming that the diffuse gas
uniformly fills the $32''$ spherical extraction volume, we 
use the normalization of the thermal component to find an 
electron density $<n_e^2>^{1/2} = 0.016$\cmc, corresponding to a
total hot gas mass of $3 \times 10^7\Ms$.

The presence of a hard non-thermal point source at the optical center 
of the galaxy coincident with a FIRST radio source is suggestive of a 
low luminosity AGN. However, emission from the region outside a $3''$ 
circle excluding the nucleus is still not well represented by a
thermal component alone, indicating the presence of unresolved
stellar sources in addition to the nuclear source.
 This is consistent with the optical classification by 
Ho \etal (1997a, 1997b) of  NGC 4435 as a transition object, 
i.e. a galaxy with nuclear emission line properties intermediate 
between an AGN and an HII region. They proposed that these line
ratios were naturally explained by emission from a nuclear AGN, 
contaminated by emission from nearby HII regions. 
Such recent star formation activity could 
increase the heavy element abundance in the surrounding diffuse gas to
near solar values.
We thus repeated the spectral analysis with $A$ fixed at $1.0\,\Zs$ 
(and Galactic absorption)
to determine how sensitive our results are to this parameter.
We found that the temperature, photon index and hard band flux 
were unchanged; while the total soft band flux decreased by 
$\sim 8\%$. However, the normalization of the thermal component 
decreased with increasing abundance by a factor of $8.5$ between
$A=0.1\,\Zs$ and $1.0\,\Zs$ reducing our estimates of the
electron density and hot gas mass each by as much as a factor $\sim 3$.
Thus, for Galactic absorption, 
$M \sim 3 \times 10^7\Ms$ should be viewed as an upper limit on the
X-ray emitting gas mass in NGC 4435.
Combined with estimates of the atomic and molecular gas (each 
$\lesssim 4 \times 10^7\Ms$, Kenney \etal 1995), this gives a 
total gas mass in NGC 4435 of $\lesssim 10^8\Ms$. 

We also checked our sensitivity to the chosen Galactic absorption. 
Ho \etal (1997b) measured an intrinsic color index $E(B-V) = 0.68$ for
the inner region of NGC 4435 using the $H\alpha/H\beta$ line ratios. 
Assuming a normal dust to gas ratio 
(Equation \ref{eq:dust}), this would imply a total (intrinsic plus Galactic)
hydrogen absorption column $n_{\rm H} = 3.85 \times
10^{21}$\cms. Fixing the absorbing column at this 
optically motivated value  
(for $A=0.1\,\Zs$ and $1.0\,\Zs$), we find the temperatures 
($kT =0.18^{+0.06}_{-0.04}$ and $kT =0.18^{+0.02}_{-0.01}$) and photon
indices ($\Gamma=1.38^{+0.21}_{-0.36}$ and
$\Gamma=1.47^{+0.28}_{-0.31}$), respectively, agree well with our 
previous fits, 
using only Galactic absorption, with no improvement in $\chi^2$.
There is little change in the hard ($< 10\%$ decrease) and soft
($15-30\%$ increase) intrinsic luminosities for the non-thermal
component because its spectral parameters are largely fixed by the 
hard emission that is unaffected by the increased absorption. 
The most significant increase is, as expected, in the
normalization and inferred absorption corrected luminosity of the 
thermal component.  The soft band intrinsic luminosity and
normalization increased by factors of $\sim 15 - 20$, thus 
increasing the inferred electron density ( and upper limit on the gas mass) 
by factors $\sim 4.6$. These measured values are still 
consistent with the gas mass and density estimates by Kenney \etal
(1995).

In Figure \ref{fig:n4435spec} we show the spectrum and two extremes of
these fits. In the left panel we show the best fit absorbed power law 
+ MEKAL model with fixed
$A=0.1\,\Zs$ and Galactic absorption only; while in the right panel we
show the best fit of the same model with fixed 
solar abundances and absorption $n_{\rm H}=3.85 \times 10^{21}$\cms inferred 
from the measurements of optical extinction. The reader should note
that the effects of increasing the abundance 
and also increasing the intrinsic absorption act oppositely on the 
electron density and hence the gas mass. For example, for the extreme
case of solar abundances and high intrinsic absorption, the electron
density ($n_e \sim 0.025$\cmc) and total X-ray gas mass 
($M \lesssim 4.7 \times 10^7\Ms$) are only a factor of $1.6$ greater 
than that in the low abundance model assuming only Galactic absorption.


\begin{figure}[t]
\epsscale{0.5}
\epsfig{file=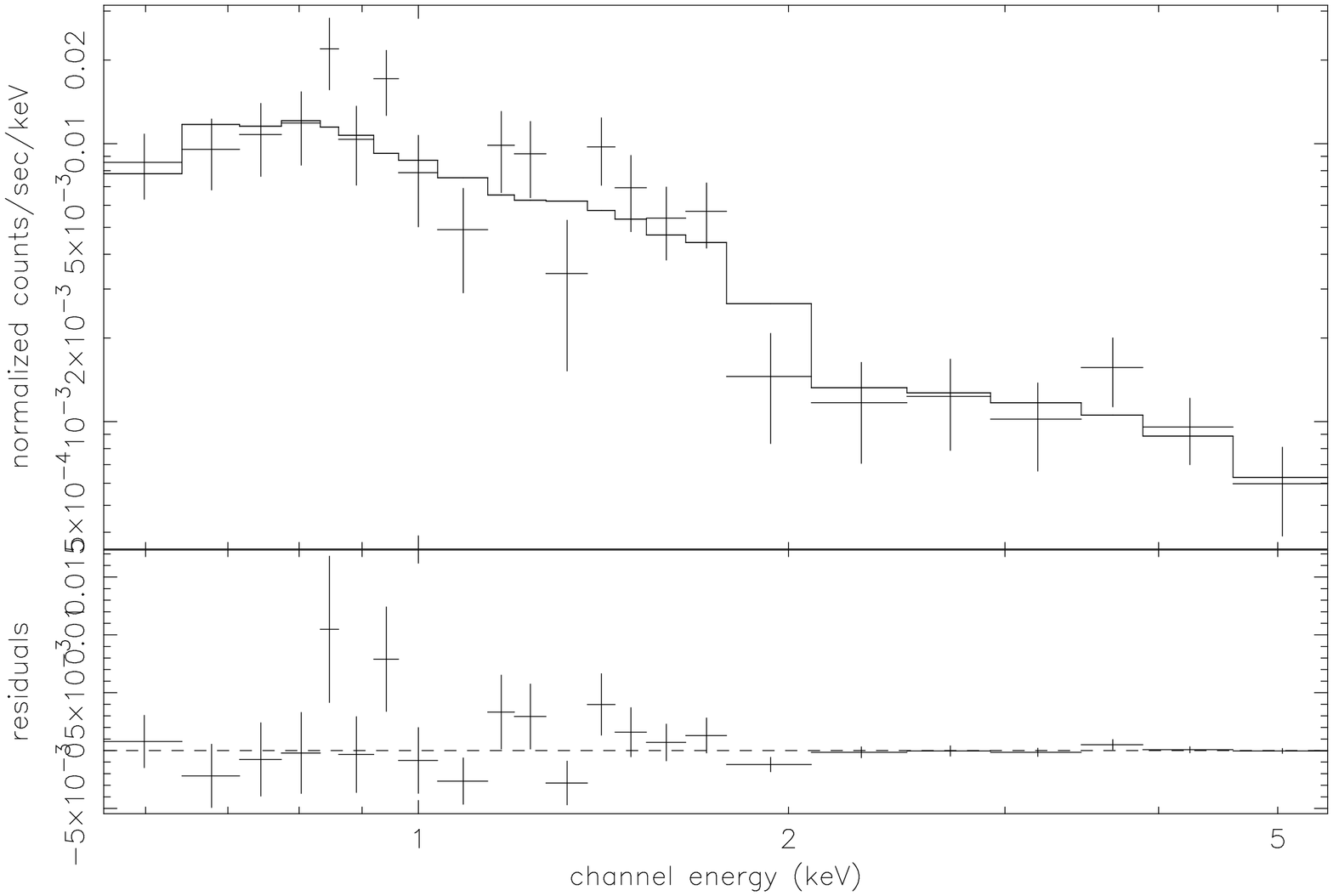,height=3in,width=3in,angle=270}
\hspace{0.3cm}\epsfig{file=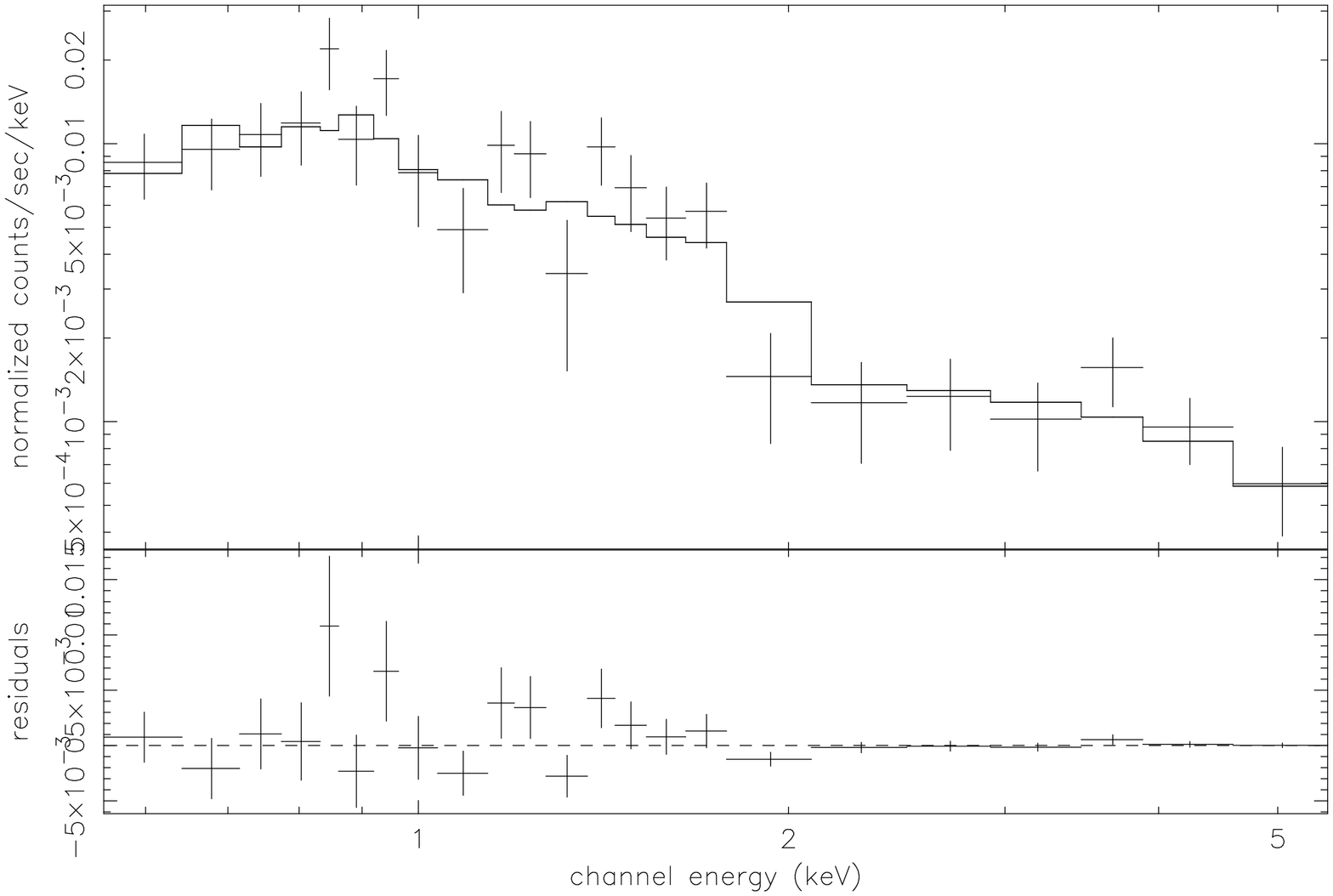,height=3in,width=3in,angle=270}
\caption{ The X-ray spectrum and absorbed power law + MEKAL model fits
for NGC 4435 for fixed
(left) low abundance $0.1\,\Zs$ and Galactic absorption
($kT=0.27^{+0.09}_{-0.07}$\,keV and $\Gamma=1.22^{+0.26}_{-0.25}$) 
and (right) solar
abundance with moderate absorption, 
$n_{\rm H}= 3.85 \times 10^{21}$\cms, inferred from optical extinction 
measurements 
($kT=0.18^{+0.02}_{-0.01}$\,keV and $\Gamma=1.47^{+0.28}_{-0.31}$).
}
\label{fig:n4435spec}
\end{figure}


The X-ray properties of NGC 4435 are also qualitatively explained by
a collision with NGC 4438. Little X-ray emission is found in
the outer regions of the optical halo in NGC 4435, but rather weak 
emission is
concentrated in the inner nuclear regions with the peak soft emission, from
gas and/or recent star formation, displaced from the peak hard emission, 
possibly from a low luminosity AGN, located at the galaxy's optical
center. A collision with NGC 4438 would strip gas more easily from the 
outer regions of the smaller of the two galaxy collision partners, 
thus accounting for the absence of X-ray emission in the outer optical
halo of NGC 4435. In galaxy collisions the gas distribution in the
inner disk of the galaxy is driven toward its nucleus and distorted by
the collision such that star formation is often induced offset from
the galactic center. This could account for the high density
($4000$\cmc) of $10^4$\,K gas and younger than normal stellar distribution
observed by Kenney \etal (1995) in the nuclear region of NGC 4435,  
its transition object classification, and the offset of the soft and
hard X-ray emission peaks. Further in this scenario one
would expect the pre-collision gas mass and densities in the outer
disk of NGC 4435 to have been much higher than presently
observed. Even given the lower limit on the densities observed today
in NGC 4435 ($n_e \gtrsim 0.01$\cmc), ram pressure due to the ISM-ISM
interactions in a collision with NGC 4438 could dominate over ICM-ISM
interations. The ram pressure on gas in NGC 4438 due to 
ISM-ISM interactions in a high velocity ($\sim900$\kms) 
galaxy-galaxy collision with NGC 4435 would exceed the 
the ram pressure on gas in NGC 4438 due to its motion through
the Virgo ICM by at least a factor of $10$.

\subsection{Point Sources}
\label{sec:points}

We identified point source candidates in the $6'.6 \times 9'.4$ field shown
in Figure \ref{fig:bigmap} using a multiscale 
wavelet decomposition algorithm in three energy bands: broadband 
($0.3-10$\,keV), soft band ($0.3-2$\,keV) and hard band ($2-10$\,kev).
The wavelet decomposition threshold was set at $5$ photons. 
In addition to the northwestern and southeastern outflow regions in 
NGC 4438 and the central region of NGC 4435, we 
detected $30$ X-ray point sources in the broadband. 
These are listed in Table \ref{tab:points} and shown in Figure 
\ref{fig:bigmap}.
Backgrounds were determined by  local annuli around each source. We consider 
$10$ source counts necessary for a flux measurement in a given bandpass. 
Assuming an effective photon index of $1.4$ 
for a power law model with Galactic absorption
($2.64 \times 10^{20}$\cms), $10$ source counts 
correspond to an absorbed flux limit in the $0.3-10$\,keV band of 
$3.6 \times 10^{-15}$\ergscm and limiting luminosity at $16.1$\,Mpc of 
$1.2 \times 10^{38}$\ergs.

\clearpage
\begin{deluxetable}{cccc}
\tabletypesize{\small}
\tablewidth{0pc}
\tablecaption{Broadband Point Sources\label{tab:points}}
\tablehead{
\colhead{RA (J2000)}  & \colhead{Dec (J2000)} & \colhead{Net Source} &
\colhead{Source Radius} \\
\colhead{(hh mm ss)}   & \colhead{(deg arcmin arcsec)} &
\colhead{Counts} & \colhead{arcsec}  \\
}
\startdata
$12$ $27$ $29.177$ & $+13$ $03$ $56.91$  &$8.4 \pm 3.4$  & $3.94$ \\
$12$ $27$ $36.757$ & $+12$ $58$ $39.59$  &$11.1 \pm 3.6$  & $1.97$ \\
$12$ $27$ $37.123$ & $+13$ $04$ $40.23$  &$14.9 \pm 4.5$  & $4.92$ \\
$12$ $27$ $37.193$ & $+13$ $01$ $21.46$  &$13.1 \pm 3.9$  & $2.46$ \\
$12$ $27$ $37.665$ & $+13$ $00$ $36.20$  &$15.7 \pm 4.4$  & $2.96$  \\
$12$ $27$ $37.999$ & $+13$ $04$ $51.06$  &$22.7 \pm 5.5$  & $4.92$  \\
$12$ $27$ $38.202$ & $+13$ $01$ $58.86$  &$9.2 \pm 3.5$  & $2.45$ \\
$12$ $27$ $40.423$ & $+13$ $04$ $56.97$  &$17.7 \pm 4.5$  & $3.94$ \\
$12$ $27$ $40.962$ & $+13$ $04$ $55.98$  &$21.7 \pm 4.8$  & $2.95$ \\
$12$ $27$ $41.503$ & $+12$ $58$ $20.91$  &$15.6 \pm 4.1$  & $2.46$ \\
$12$ $27$ $42.748$ & $+13$ $00$ $02.26$  &$47.9 \pm 7.1$  & $1.97$  \\
$12$ $27$ $42.916$ & $+13$ $02$ $35.76$  &$5.0 \pm 2.7$  & $2.46$  \\
$12$ $27$ $43.286$ & $+12$ $59$ $58.32$  &$13.0 \pm 4.0$  & $1.96$  \\
$12$ $27$ $43.657$ & $+13$ $00$ $25.8$   &$7.8 \pm 3.0$  & $1.47$  \\
$12$ $27$ $44.196$ & $+12$ $58$ $33.21$  &$12.9 \pm 4.1$  & $3.45$  \\
$12$ $27$ $44.633$ & $+13$ $00$ $25.39$  &$22.6 \pm 5.1$  & $1.96$  \\
$12$ $27$ $45.138$ & $+12$ $59$ $43.07$  &$6.4 \pm 2.9$  & $1.97$  \\
$12$ $27$ $45.205$ & $+13$ $00$ $19.48$  &$8.1 \pm 3.2$  & $1.48$  \\
$12$ $27$ $45.239$ & $+12$ $59$ $12.57$  &$20.1 \pm 4.7$  & $2.95$  \\
$12$ $27$ $45.575$ & $+13$ $00$ $42.61$  &$22.3 \pm 5.8$  & $1.96$  \\
$12$ $27$ $46.485$ & $+12$ $59$ $29.79$  &$27.6 \pm 5.4$  & $1.97$  \\
$12$ $27$ $46.586$ & $+12$ $59$ $07.65$  &$6.2 \pm 2.7$  & $1.48$  \\
$12$ $27$ $47.125$ & $+13$ $00$ $21.94$  &$13.4 \pm 4.0$  & $1.97$  \\
$12$ $27$ $47.360$ & $+13$ $00$ $45.56$  &$7.8 \pm 3.0$  & $1.48$  \\
$12$ $27$ $47.360$ & $+13$ $01$ $01.30$  &$27.6 \pm 5.8$  & $2.46$   \\
$12$ $27$ $47.394$ & $+13$ $00$ $53.92$  &$7.0 \pm 2.9$  & $1.48$  \\
$12$ $27$ $47.427$ & $+12$ $59$ $14.53$  &$32.1 \pm 5.8$  & $1.97$  \\
$12$ $27$ $50.391$ & $+13$ $04$ $14.66$  &$4.3 \pm 2.2$  & $1.97$  \\
$12$ $27$ $54.362$ & $+13$ $01$ $18.51$  &$13.9 \pm 5.3$  & $5.91$  \\
$12$ $27$ $54.666$ & $+13$ $03$ $18.56$  &$29.9 \pm 5.6$  & $2.95$   \\
\enddata
\tablecomments{ The broadband energy bandpass is $0.3 - 10$\,keV. 
Source radii denote the radius of the circular region used
to exclude the point source from the spectral analyses of the diffuse gas.}
\end{deluxetable}
\clearpage

The sources in NGC 4438 are strongly clustered. 
Most sources lie to the south of the galactic
center and either extend  along the line of a bright extended X-ray filament 
or lie just east of the stellar tidal ridge along the leading 
(southeastern) edge (Kenney \etal 1995) of the galaxy.   If we consider a 
$2'$ circular region around the telescope aim point encompassing NGC 4438, 
where we expect our sample of sources with $\gtrsim 10$ photon counts to 
be complete, we find $8$ independent sources, i.e. $6$ sources in the soft
band only (red), $1$ source in the hard band only (cyan), and $1$ source that
exceeds $10$ counts in both (magenta).   Using the 
Cosmic X-ray Background (CXB) cumulative 
number counts curves $N(>S)$ from the Chandra Deep Field North $1$\,Ms survey
(Brandt \etal 2001) combined with our flux detection
limits, we would expect to see $3 \pm 1$ background sources in the 
soft band and $1^{+1.0}_{-0.5}$ background source in the hard band. 
Since NGC 4438
is known to have a significant population of old stars (Bonatto \etal
2000), we expect that most of the excess sources in NGC 4438 
are low mass X-ray binaries (LMXB's). Using the $N(>S)$ curve for LMXB's 
in the Milky Way (Grimm \etal 2002) scaled by the ratio of the dynamical 
masses of the galaxies, we expect $4.8^{+0.2}_{-0.6}$ LMXB's in NGC 4438 above
$10$ counts and we detect $4$,  consistent with this prediction. The 
flux corresponding to $10$ counts in the hard band is, however, above
the high luminosity cut-off of $2.7 \times 10^{38}$\ergs for LMXB's given
by those authors.  The single hard source detected above the CXB 
prediction for this band may, if not part of the cosmic X-ray
background, be a high mass
X-ray binary (HMXB). Since the cumulative number distribution of HMXB's is 
highly correlated with the current star formation rate (Grimm \etal 2003), we
can use the hard band data to place an upper limit on the current star 
formation rate ($sfr$) in NGC 4438 outside of the nucleus
of $sfr \lesssim 0.40^{+0.24}_{-0.13}\Ms$\,yr$^{-1}$. This is consistent
with an estimate of the upper limit on star formation determined from 
FIR data, $sfr \lesssim 0.42^{+0.54}_{-0.24}\Ms$\,yr$^{-1}$, where we
have used
$\log(L_{\rm FIR}/L_\odot) = 9.38 \pm 0.36$ from Bonatti \etal (2000) and 
the star formation rate - FIR luminosity correlation 
from Grimm \etal (2002).
The lack of point sources 
to the north of the bulge is also understandable in the context of a 
collision between NGC 4438 and NGC 4435. Since the strength of
galaxy-galaxy tidal interactions vary significantly over galactic
scales, tidal interactions from such a
collision would disrupt existing stellar distributions more strongly
to the north, near the point of closest approach, than to the south or
in the tightly bound bulge. 

Sources are also seen in Figure \ref{fig:bigmap} to be clustered near 
NGC 4435. Four source candidates, in addition to the nucleus and soft 
X-ray peak $4.''1$ to its northeast discussed previously, are
identified by the algorithm. While these sources are consistent with 
being point sources, the broadening of the point spread function 
this far from the telescope aim point makes it difficult to distinguish 
between point sources and regions of extended emission that 
may be associated with the galaxy.

We find no evidence for the weak (marginally
significant) X-ray emission found in Einstein HRI maps (Kotanyi \etal
1983; Fabianno \etal 1992) located 
roughly midway between NGC 4438 and NGC 4435 near a peak in HI
emission (Hibbard \& van Gorkam 1990). Using backgrounds determined locally 
from several positions on the detector chips away from the galactic emission 
to subtract the expected Virgo cluster emission, we find that
the excess X-ray flux in the $0.3-2$\,keV energy band from a circular 
region of radius $30''$ centered at
($12^h27^m45.9^s$,$+13^\circ3'2''.2$), 
where the  marginal Einstein detection was reported,
is consistent with zero in all cases. Thus assuming a thermal 
spectral model with $kT=0.34$\,keV and Galactic absorption, 
we place a $3\sigma$ upper limit
on the excess flux in this region of  
$\lesssim 3.7 \times 10^{-15}$\ergscm.
However, it should be noted that
this region crosses the chip gap between S2 and S3 in our observation,
so a compact source falling within the chip gap might have been missed.


\section{Conclusions}
\label{sec:conclude}

In this paper we presented results from a $25$~ks Chandra ACIS-S observation 
of the peculiar spiral galaxy NGC~4438  in the Virgo Cluster  taken 
on 29 January 2002. We
were able to separately study the diffuse extended emission
features and the central $1$\,kpc surrounding the nuclear
source. We found a remarkable correspondence between the X-ray emission
features in both regions and those seen in H$\alpha$ images. We also 
observed X-ray emission from the nearby galaxy NGC 4435 and identified
point sources in both galaxies.

In summary we found:
\begin{enumerate}

\item{The extended emission in NGC 4438 can be separated 
into  a bulge, within a 
radius $\sim 2.3$\,kpc from the galaxy's center, and a 
network of morphologically disturbed filaments with thicknesses 
$\sim 1$\,kpc and lengths $4-10$\,kpc extending out of the plane of the 
disk west and southwest of the bulge. These
X-ray filaments correspond closely to similar 
features observed in H$\alpha+$[N II] (Kenney \etal 1995; 
Keel \& Wehrle 1993; Chemin \etal 2003) with low gas velocities 
($\lesssim 200$\kms) such that the gas may still be
gravitationally bound to the galaxy.}

\item{The extended X-ray emitting gas associated with NGC 4438
is well fit by a single temperature MEKAL model with
fixed Galactic absorption ($n_H=2.64 \times 10^{20}$\cms), temperature
$kT=0.42^{+0.07}_{-0.04}$\,keV, and element abundances 
$A=0.11^{+0.05}_{-0.03}\,\Zs$.
 The observed $0.3-2$\,keV flux
for the extended gas is $3.35 \times 10^{-13}$\ergscm corresponding to
an unabsorbed X-ray luminosity at $16.1$\,Mpc of 
$1.24 \times 10^{40}$\ergs.
We find no significant X-ray spectral
differences between gas in the filaments and gas in the bulge.}
 
\item{Assuming simple geometries for the extended features and uniform
filling of their associated volumes, we infer electron densities
$<n_e^2>^{1/2} \sim 0.02 - 0.04$\cmc
for both gas in the filaments and the bulge, and cooling times
$t_{cool} \sim 4 - 7 \times 10^8$\,yr.
The total mass of hot gas in a $95''$ aperture surrounding the galaxy is
$M \lesssim 3.7 \times 10^{8}\Ms$. 
}

\item{The low temperatures ($0.4$\,keV or $4.6 \times 10^6$\,K) and low
velocities observed in the extended features, as well as the
asymmetrically disturbed morphology and strong correlation between 
H$\alpha$ and X-ray emitting gas, are most naturally explained by tidal
interactions and multiphase ISM-ISM interactions, acting during a 
galaxy-galaxy 
collision that occurred $\sim 10^8$ years ago, as NGC 4438 passed 
$\sim 5$\,kpc to the south of NGC 4435's center with relative velocity  
$\sim 900$\kms.
}

\item{X-ray emission in the center of NGC 4438 is observed
from  two regions corresponding to the two outflow shells observed in 
H$\alpha+$[N II] with HST by Kenney \& Yale (2002) and to the two peaks 
observed in the radio continuum by Hummel \& Saikia (1991). The
brighter feature extends $\sim 360$\,pc to the northwest of the 
optical nucleus with the peak of the 
X-ray emission $\lesssim 0''.7$ from the compact nuclear source. The 
fainter southeastern outflow shell consists of patchy,
shell-like emission in a circular aperture of radius $\sim 130$\,pc
located $730$\,pc southeast of the nuclear source. The X-ray spectrum
of the northwest  outflow 
region, including the compact nuclear source, is well
fit by an absorbed MEKAL model thermal component with temperature 
$kT = 0.58^{+0.04}_{-0.10}$\,keV, abundance $0.14\,\Zs$, and 
absorbing column $n_H = 1.9^{+0.10}_{-0.04} \times 10^{21}$\cms 
 and a heavily obscured 
($n_H = 2.9^{+3.1}_{-2.0} \times 10^{22}$\cms) AGN-like (photon
index fixed at $2.0$) nuclear source.}

\item{The $0.3-2$\,keV ($2-10$\,keV) unabsorbed X-ray luminosities from
the northwest outflow region are 
$1.02 \times 10^{40}$\ergs ($2.6 \times 10^{38}$\ergs) for the
thermal component and 
$2.4 \times 10^{39}$\ergs ($2.0 \times 10^{39}$\ergs) for the 
non-thermal component, with large uncertainties in the latter due to 
uncertainties in the absorption correction.  
Assuming the same spectral model for
the gas in the southeastern outflow bubble, the inferred electron
densities and cooling times in both outflow
bubbles are $0.5$\cmc and 
$3 \times 10^{7}$\,yr.
The total amount of hot gas in the outflow regions is  
$\sim 3.6 \times 10^6\Ms$. The spectral properties and extent of
the outflow features are consistent with gas heated by shocks driven
into the surrounding ISM with shock velocities $\gtrsim 600$\kms by 
nuclear outbursts $1-2 \times 10^6$ years ago. 
The large amount of kinetic energy carried in the outflow gas 
($\gtrsim 3.5 \times 10^{54}$\,ergs for the northwestern outflow bubble),
coupled with low upper limits ($\lesssim 0.1\Ms$\,yr$^{-1}$) on
recent star formation in the nucleus of NGC 4438, favor
an obscured AGN, over a compact starburst, as the nuclear source.} 

\item{Asymmetric X-ray emission is observed in a $2.5$\,kpc circular 
aperture about the optical center of NGC 4435 with soft ($0.3-2$\,keV)
and hard ($2 - 10$\,keV) band total observed fluxes of 
$7.0 \times 10^{-14}$\ergscm and $1.25 \times 10^{-13}$\ergscm. 
The peak of the hard emission occurs at the optical center of the 
galaxy coincident with a FIRST radio source; while the 
peak of the soft emission is displaced $316$\,pc to the northeast. 
The spectrum is well represented by an absorbed two component model
consisting of a non-thermal power law component, most probably from 
both a low luminosity AGN at the nucleus and unresolved stellar sources, 
 and a $0.2-0.3$\,kev thermal MEKAL plasma component from diffuse ISM gas.
For fixed $0.1\,\Zs$ abundance and
Galactic absorption, we infer an electron density $n_e \sim 0.016$\cmc and,
assuming a uniform filling factor, an X-ray gas mass 
$M \sim 3 \times 10^7\,\Ms$. However, the electron density 
and X-ray emitting gas mass are sensitive to the 
assumed element abundance (decreasing by a factor 
$\sim 3$, as $A$ is varied from $0.1-1.0\,\Zs$) and 
absorbing column (increasing by a factor of $4.6$ as 
$N_H$ varies from Galactic to $3.85 \times 10^{21}$\cms, motivated by 
optical extinction measurements) and thus are highly uncertain. 
The fact that gas is primarily concentrated near the nucleus is 
consistent with much of the outer gas having been 
stripped from NGC~4435, the smaller of the two galaxies, 
during a collision with NGC~4438.
}

\item{Point sources, identified above the broadband ($0.3-10$\,keV)
flux (luminosity) limit of $3.6 \times 10^{-15}$\ergscm 
($1.2 \times 10^{38}$\ergs), 
are found asymmetrically clustered mainly 
to the south of the galactic center in NGC 4438 consistent with the 
stars north of the galactic center having been disrupted in a 
collision with NGC 4435. Excess sources above the Cosmic X-ray 
Background prediction in the soft band 
are consistent with the expected number of LMXB's based on the mass of
the galaxy. The  single hard band
source places an upper limit on the current star formation rate 
outside the nucleus of 
$\lesssim 0.4^{+0.24}_{-0.13}\Ms$\,yr$^{-1}$. 
}

\end{enumerate}


\acknowledgements

 This work has been supported in part by NASA contract NAS8-39073, and
  the Smithsonian Institution. MEM also acknowledges support  
  from the Radcliffe Institute for Advanced Study at Harvard University. 
 This work has made use of the NASA/IPAC Extragalactic Database (NED)
 which is operated by the Jet Propulsion Laboratory, California
 Institute of Technology,  under contract with the National
 Aeronautics and Space Administration. We wish to thank Sebastian
  Heinz, Ralph Kraft, Paul Nulsen, and Laurent Chemin for useful discussions.
 

\end{document}